%% file: AircraftStateGen.tex
\documentclass[twocolumn,journal]{IEEEtran}
\usepackage[noadjust]{cite}
\usepackage[T1]{fontenc}
\usepackage[latin9]{inputenc}
\usepackage{array}
\usepackage{verbatim}
\usepackage{float}
\usepackage{enumitem}
\usepackage{amsmath}
\usepackage{amssymb}
\usepackage{graphicx}
\usepackage{multirow}
\usepackage[unicode=true,
 bookmarks=true,bookmarksnumbered=true,bookmarksopen=true,bookmarksopenlevel=1,
 breaklinks=false,pdfborder={0 0 0},pdfborderstyle={},backref=false,colorlinks=false]
 {hyperref}
\hypersetup{pdftitle={Your Title},
 pdfauthor={Your Name},
 pdfpagelayout=OneColumn, pdfnewwindow=true, pdfstartview=XYZ, plainpages=false}
 
\makeatletter

\providecommand{\tabularnewline}{\\}
\floatstyle{ruled}
\newfloat{algorithm}{tbp}{loa}
\providecommand{\algorithmname}{Algorithm}
\floatname{algorithm}{\protect\algorithmname}

\usepackage[caption=false,font=footnotesize]{subfig}
\usepackage{tikz}
\usepackage{algorithm}% http://ctan.org/pkg/algorithms
\usepackage[noend]{algpseudocode}

\makeatother

\begin{document}
\title{Analytical Aircraft State and IMU Signal Generator from Smoothed Reference Trajectory}
\author{Mr. Austin Costley, Dr. Randall Christensen, Dr. Robert C. Leishman, Dr. Greg Droge
\thanks{This work was supported by Air Force Research Laboratory, Wright-Patterson
Air Force Base, OH.\protect \\
A. Costley is with the Electrical and Computer Engineering Department,
Utah State University, Logan, UT 84322 USA (e-mail: austin.costley@usu.edu)\protect \\
R. Christensen is with the Electrical and Computer Engineering Department,
Utah State University, Logan, UT 84322 USA (e-mail: randall.christensen@usu.edu)\protect \\
R. Leishman is with the ANT Center, Air Force Institute of Technology,
Wright-Patterson Air Force Base, OH 45433 USA (e-mail: robert.leishman@afit.edu)\protect \\
G. Droge is with the Electrical and Computer Engineering Department,
Utah State University, Logan, UT 84322 USA (e-mail: greg.droge@usu.edu)}}
\maketitle
\begin{abstract}
This work presents a method for generating position, attitude, and velocity states for an aircraft following a smoothed reference trajectory. The method also generates accelerometer and gyro measurements consistent with the aircraft states. This work describes three corner smoothing algorithms for generating a smoothed reference trajectory from a series waypoints that accounts for limitations in the physical system. The smoothed reference trajectory, and curvilinear motion theory are used to generate the lower-order states. A coordinated turn maneuver is then applied to generate accelerometer and gyro measurement estimates.
\end{abstract}

\section{Introduction}
\input{sections/Introduction.tex}

\section{Background\label{sec:Background}}
\input{sections/Background.tex}

\section{Corner Smoothing\label{sec:Path-Smoothing}}
\input{sections/PathSmoothing.tex}

\section{IMU Signal Generation\label{sec:State-Generation}}
\input{sections/ImuSignalGeneration.tex}

%\section{Covariance Propagation\label{sec:Covariance-Propagation}}
%\input{sections/CovariancePropagation.tex}

\section{Results\label{sec:Results}}
\input{sections/Results.tex}

\section{Conclusion}
\input{sections/Conclusion.tex}

\bibliographystyle{IEEEtran}
\bibliography{IMU-signal-generation-paper}
% \bibliography{newbib.bib}
\end{document}

%% file: sections/Introduction.tex
An objective for the mission and path planning of unmanned aerial vehicles (UAVs) is to determine a flyable path that meets mission requirements. The characteristics of a flyable path vary based on the dynamics of the target platform. For example, a flyable path for a fixed wing aircraft must have continuous course angle and curvature and have constraints on the maximum curvature and curvature rate. Additional mission requirements vary based on the application, but may include entering areas of interest, low probability of collision or detection, or mission duration. This work provides a method for generating aircraft states and inertial measurement unit (IMU) measurements along a flyable reference trajectory to evaluate performance against mission requirements. 

This work assumes that a piecewise linear candidate path (or series of waypoints) has been generated. However, a piecewise linear path is not flyable for a fixed-wing UAV due to the discontinuities in course angle. A path smoothing algorithm \cite{yang_analytical_2010, ravankar_path_2018} can be used to convert the piecewise linear candidate path into a smoothed reference trajectory. While many approaches to path smoothing exist, the method presented in this work results in a series of path segments that have analytical solutions for position, course angle, and curvature.

The segment types used in this work are lines, arcs, and transition segments. The transition segments are used to connect the lines and arcs and maintain continuity in course angle and curvature. One curve used as a transition segment is the clothoid, which provides a linearly changing curvature per unit length \cite{scheuer_continuous-curvature_1997}. A drawback to the clothoid curve is the dependence on the Fresnel Integrals that do not have a closed form solution, however, accurate approximations exist and can be used to simplify the computation \cite{vazquez-mendez_clothoid_2016}. In addition to the clohtoid, Lekkas et al. \cite{lekkas_continuous-curvature_2013} present a method for path smoothing using a Fermat Spiral which avoids the costly computation of the Fresnel Integrals while maintaining continuous curvature. This work presents an algorithm for generating 3 types of smoothed paths (Arc Fillet, Clothoid Fillet, and Fermat Fillet) parameterized by a maximum curvature and maximum curvature rate. 

An important aspect of mission and path planning for UAVs is to account for uncertainty during the operation. Uncertainties exist in three components in planning, namely, vehicle position and orientation, environmental, and vehicle motion uncertainties. Path planning techniques often model some or all of these uncertainties to determine an optimal path \cite{blackmore_chance-constrained_2011, du_toit_robot_2012}. In this work, a fixed-wing UAV with an inertial navigation system (INS) \cite{groves_principles_2013} is considered. The propagation of the state estimates and uncertainty of an INS is dependent on measurements from a strapdown IMU \cite{farrell_aided_2008}. Thus, a path planning algorithm considering the uncertainty of such a system would require estimating aircraft states and the associated IMU measurements along the candidate path. 

A common application for IMU measurement generation is for trajectory reconstruction. Savage \cite{savage_strapdown_2000} presents a high-fidelity trajectory reconstruction method that generates nearly perfect IMU signals. To achieve that level of accuracy, the method uses complicated equations that are evaluated at very small time steps. For path planning applications, the fidelity of the resulting IMU measurement estimates can be relatively low and more emphasis is placed on the computational efficiency. This work presents an approach to using the path geometry of a smoothed reference trajectory to analytically generate aircraft states (position, orientation) and IMU measurements (specific force and angular rates) for a UAV during straight-and-level flight and during a coordinated turn. The resulting IMU measurements are consistent over large sampling times.

This work presents a method, ASG, for generating aircraft states and IMU measurements for a smoothed reference trajectory. The method presented provides a description of three corner smoothing algorithms in a unified framework. The corner smoothing algorithm provides a series of path segments that form a smoothed reference trajectory. The segment geometry is leveraged to generate aircraft states along the trajectory. The final stage of the ASG method is a novel approach to analytically generating IMU signals using the path geometry and a coordinated turn motion model.

The next section describes background information that is used throughout this work. Section \ref{sec:Path-Smoothing} presents the corner smoothing algorithm that converts a series of waypoints to a path of continuous course angle. The path segment definitions (see Section \ref{sec:background_segment_defs}) are used to determine the position and course angle of the aircraft. Section \ref{sec:State-Generation} describes a method for determining the roll, pitch, body frame accelerations, and angular rates of the aircraft assuming straight and level flight during straight line segments, and a coordinated turn for transition segments and arcs. Finally, Section \ref{sec:Results} provides results for the ASG method that show the resulting states and that the accuracy of the IMU measurements is maintained for large time steps.

%% file: sections/Background.tex
This section presents background information used throughout this work. Section \ref{sec:background_notation} describes the notation. Section \ref{sec:background_segment_defs} defines the segment types used in the corner smoothing algorithm. Accelerations and angular rates along the smoothed path are determined by curvilinear motion theory which is discussed in Section \ref{sec:background_curv}.  Finally, Section \ref{subsec:RII} describes a method for calculating accelerations and angular rates using INS kinematic equations.

%%%%%%%%%%
% Notation
%%%%%%%%%%
\subsection{Notation \label{sec:background_notation}}
In this work, curvature rate refers to the derivative of curvature with
respect to path length, $s$, thus the following convention will be
used for brevity
\begin{eqnarray*}
 \frac{dk}{ds} & = & k^{\prime}
\end{eqnarray*}
when the derivative is taken with respect to time, the standard dot
notation will be used, $\dot{k}$. 

Vectors in this work will be expressed in bold face such that $\boldsymbol{v} = \left[\begin{array}{c} v_x \ v_y \ v_z \end{array}\right]^T$ for a vector in Cartesian coordinates.

This work will maintain three frames of reference in which vector quantities may be expressed, namely, inertial, $i$, body, $b$, and the Navigation or North-East-Down (NED), $n$. The frame in which a vector is expressed will be indicated by the superscript so the velocity vector in the NED frame would be expressed as $\boldsymbol{v}^n = \left[\begin{array}{c} v_n \ v_e \ v_d \end{array}\right]^T$.

The cross product matrix is used throughout this work and will be given by
\begin{equation}
\Omega = [\boldsymbol{\omega} \times] = \left[\begin{array}{ccc} 0 & -\omega_z & \omega_y \\ \omega_z & 0 & -\omega_x \\ -\omega_y & \omega_x & 0 \end{array}\right].
\end{equation}

%%%%%%%
% Corner Smoothing Segments
%%%%%%%
\subsection{Corner Smoothing Segment Definitions \label{sec:background_segment_defs}}
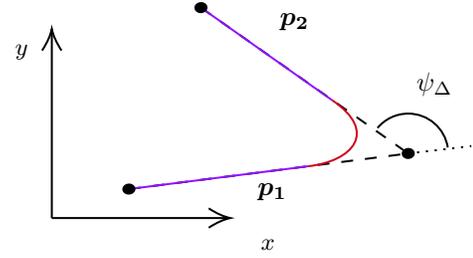
\begin{figure}
\begin{centering}
\noindent\begin{minipage}[t]{1\columnwidth}%
\begin{center}
\input{figures/path_smoothing.tex}
\par\end{center}%
\end{minipage}
\par\end{centering}
\caption{Corner smoothing example where a two segment piece-wise linear path is smoothed to create path with continuous course angle and curvature. The entry vector is shown as $\boldsymbol{p_1}$, the exit vector is shown as $\boldsymbol{p_2}$, and the course angle change is expressed as $\psi_\Delta$. \label{fig:smoothing_example}}
\end{figure}

There is a significant amount of literature on the subject of corner smoothing \cite{yang_analytical_2010, ravankar_path_2018}. As the name suggests, corner smoothing takes a corner made from the connection point of two straight path segments and generates a curved segment to transition from the first straight segment to the second straight segment. Fig. \ref{fig:smoothing_example} shows an example of a two segment piece-wise linear path with a smoothed
turn. 

The corner smoothing algorithm in this work will output a series of path segments, including straight lines, maximum curvature (minimum turn radius) arcs, and transition segments with changing curvature. The transition segments that will be used in this work are clothoid and Fermat spiral segments. For a given corner, the algorithm will generate two or three smoothing segments depending on the sharpness of the turn. A sharp turn will require three segments (transition segment, maximum curvature arc, transition segment). Whereas a gradual turn will only require two transition segments to smooth the corner. The following subsections define expressions to determine the Cartesian positions $(x,y)$, course angle $(\psi)$, and curvature $(k)$ along each of the path segment types.

\subsubsection{Line\label{subsec:line}}
The first segment type is a line which is defined by
\begin{eqnarray}
x\left(s\right) & = & x_{0}+s\cos\left(\psi_{0}\right)  \label{eq:line_x} \\
y\left(s\right) & = & y_{0}+s\sin\left(\psi_{0}\right) \label{eq:line_y} \\
\psi & = & \psi_{0} \\
k & = & 0
\end{eqnarray}
where $s$ is the length along the segment, and $\psi_{0}$ is the initial segment angle.

\subsubsection{Arc\label{subsec:Arc}}
The arc is a continuous curvature segment. In the smoothing application, the arc will be parameterized by the maximum curvature ($k_{max}$) attainable by the aircraft. The expressions that define the arc are parameterized by the polar angle, $\theta$, and are given by
\begin{eqnarray}
x\left(\theta\right) & = & x_{0}-\rho r\sin\left(\psi_{0}\right)+\rho\cos\left(\theta\right) \label{eq:arc_x} \\
y\left(\theta\right) & = & y_{0}+\rho r\cos\left(\psi_{0}\right)+\rho\sin\left(\theta\right) \label{eq:arc_eqs}\\
\psi\left(\theta\right) & = & \psi_{0}+\rho\theta\nonumber \label{eq:arc_psi}\\
k & = & \frac{\rho}{r} \label{eq:arc_k} 
\end{eqnarray}
where $r$ is the radius of the arc (the inverse of curvature), and $\rho$ indicates the direction of travel where
\begin{equation}
\rho=\begin{cases}
1, & \text{counter-clockwise (left) turn}\\
-1, & \text{clockwise (right) turn}.
\end{cases}\label{eq:direction}
\end{equation}
The polar angle, $\theta$, is related to the length on the path by the arc length formula $s=r\theta$. Thus, the arc can be sampled at a specified arc length interval by computing the associated polar angle interval and applying the equations above. Note that the convention employed here is that $\theta=0$ is associated with the point on the curve where $\psi=\frac{\pi}{2}$. 

\subsubsection{Clothoid\label{subsec:Clothoid}}
A clothoid or Euler spiral is a path segment whose curvature varies linearly by segment length \cite{levien_euler_2008}. The position along the clothoid segment is determined by computing the Fresnel Integrals given by

\begin{eqnarray}
x = C(s) = x_{0}+\int_{0}^{s}\cos(0.5\sigma_c\xi^{2}+k_{0}\xi+\psi_{0})d\xi\label{eq:fresnel_x} \\
y = S(s) = y_{0}+\int_{0}^{s}\sin(0.5\sigma_c\xi^{2}+k_{0}\xi+\psi_{0})d\xi\label{eq:fresnel_y}
\end{eqnarray}
where $s$ is the length along the clothoid segment, $\psi_{0}$ is the initial segment angle, $k_{0}$ is the initial segment curvature, $x_{0}$ and $y_{0}$ represent the starting point of the segment, and $\sigma_c$ is the curvature rate of the clothoid segment. A primary challenge when working with clothoids is that the Fresnel Integrals do not have a closed form solution so numerical solutions are required to compute positions along the path. However, there are numerous methods \cite{yang_analytical_2010, wilde_computing_2009, brambley_clothoid} for computing accurate approximations to these equations that help mitigate this drawback. 

\begin{figure}
\begin{centering}
\includegraphics[width=0.85\columnwidth]{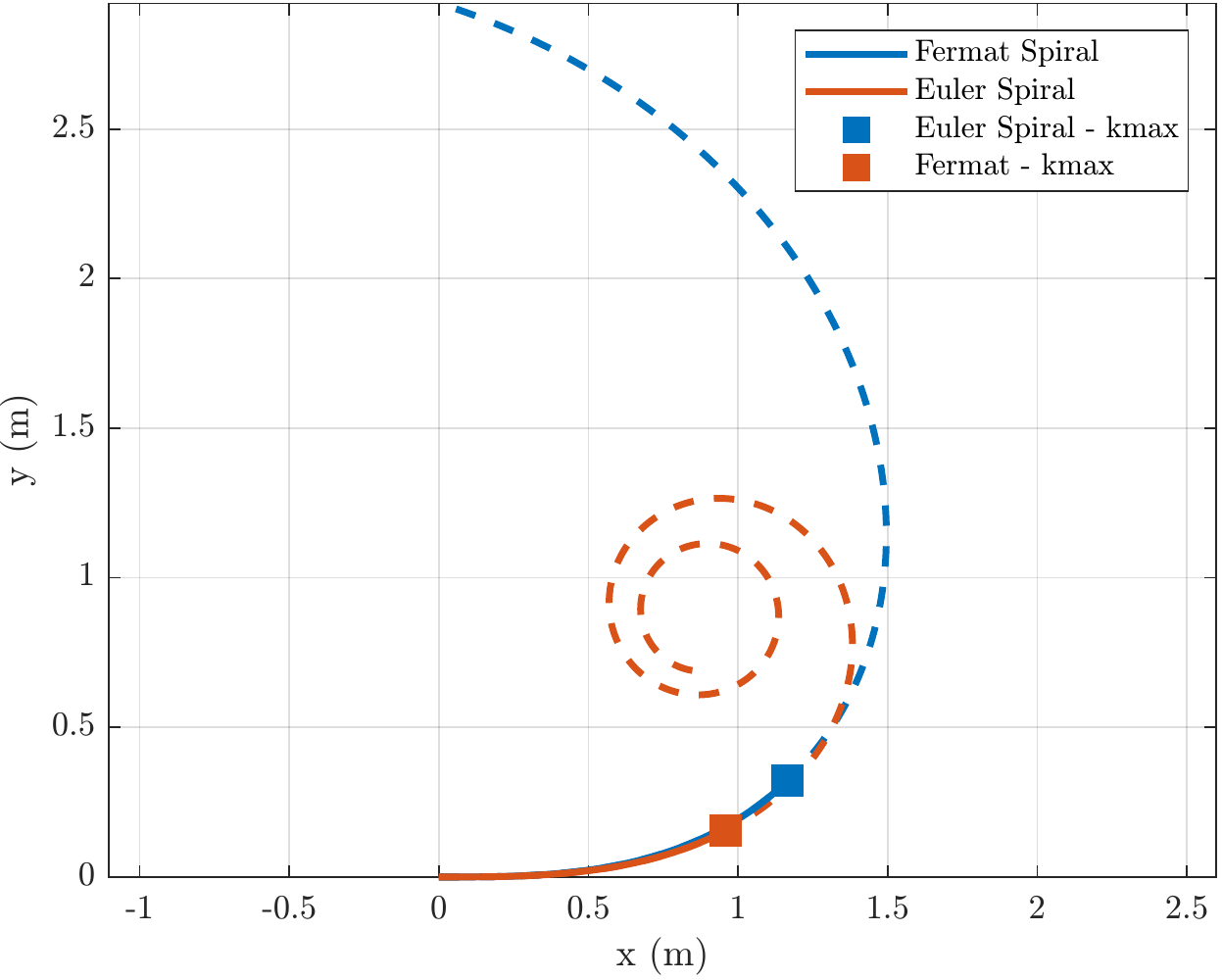}
\par\end{centering}
\caption{Euler spiral and Fermat spiral with $\psi_{0}=\frac{\pi}{2}$, $k_{0}=0$. The curvature rate for the Euler spiral is $\sigma_c=1$ and Fermat Spiral has a maximum curvature of 1 $\frac{1}{m}$. The square markers on each line indicate the location where the curvature of the segment is equal to one and the solid segments (before the markers) show the part of the segments used in the path smoothing algorithm. \label{fig:transition_segments}}
\end{figure}

The course angle of a clothoid segment changes quadratically by segment length as given by

\begin{equation}
\psi(s)=\psi_{0}+k_{0}s+0.5\sigma_cs^{2}.\label{eq:clothoid_heading}
\end{equation}
Finally, the curvature of the clothoid segment is given by

\begin{equation}
k(s)=k_{0}+\sigma_cs.\label{eq:clothoid_curvature}
\end{equation}
This shows the linear relationship between path length and curvature. An example of a Euler spiral is shown in Fig. \ref{fig:transition_segments} for a segment that is 5 meters long, has an initial segment angle and initial curvature of zero, and a curvature rate of one.

\subsubsection{Fermat Spiral\label{subsec:Fermat-Spiral}}
\begin{figure}
\begin{centering}
\includegraphics[width=0.85\columnwidth]{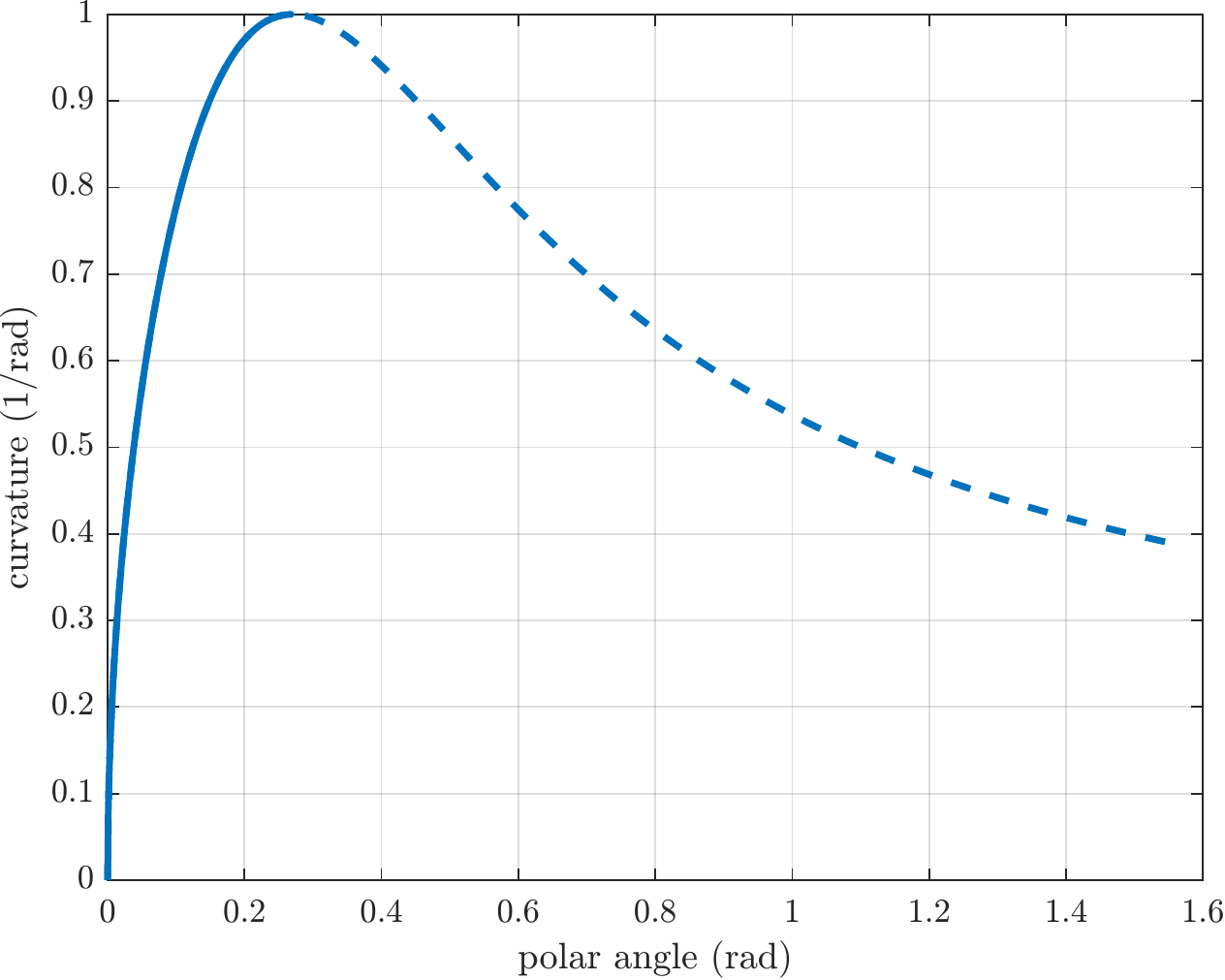}
\par\end{centering}
\centering{}\caption{Curvature of Fermat spiral for a quarter of a rotation of the polar angle with a maximum curvature of 1 $\frac{1}{m}$. The solid line shows the portion of the spiral that will be used in the transition segment. The curvature decreases after the point of maximum curvature.  \label{fig:Curvature-of-Fermat}}
\end{figure}

Another spiral that has been investigated for corner smoothing is the Fermat spiral \cite{lekkas_continuous-curvature_2013}. The Fermat spiral is an interesting corner smoothing candidate as it provides many of the benefits of the clothoid but avoids the integral computation used to compute the positions along the curve. The curve is parameterized in a polar coordinate system as $r = c\sqrt{\theta}$, where the $c$ parameter can be used to modify the curve characteristics and $\theta$ is the polar angle. Converting the curve to Cartesian coordinates results in
\begin{eqnarray}
x(\theta) & = & x_{0}+c\sqrt{\theta}\cos\left(\rho\theta+\psi_0\right) \label{eq:fermat_positionx} \\
y(\theta) & = & y_{0}+c\sqrt{\theta}\sin\left(\rho\theta+\psi_0\right). \label{eq:fermat_positiony}
\end{eqnarray}
The curve is shown in Fig. \ref{fig:transition_segments} where it can be noted that curvature does not monotonically increase as the polar angle increases, instead, the curvature increases to a maximum and then decreases until settling at a nearly constant value. The curvature is given by the non-linear equation

% \begin{figure}
% \centering{}\includegraphics[width=1\columnwidth]{figures/fermat_spiral}\caption{Fermat spiral for one complete revolution.\label{fig:Fermat-spiral}}
% \end{figure}

\begin{align}
k\left(\theta\right) & =\frac{1}{c}\frac{2\sqrt{\theta}\left(4\theta^{2}+3\right)}{\left(4\theta^{2}+1\right)^{\frac{3}{2}}}.\label{eq:fermat_curvature}
\end{align}
 Fig. \ref{fig:Curvature-of-Fermat} shows the curvature of the Fermat spiral for a quarter of a rotation of the polar angle. The figure shows that there is a single maximum for $\theta>0.$ The value of the polar angle at which the maximum curvature is achieved at

\begin{align}
\theta_{k_{max}} & =\sqrt{\frac{\sqrt{7}}{2}-\frac{5}{4}}\label{eq:theta_kmax} \approx 0.26995
\end{align}
which is independent of the shaping parameter, $c$. Beyond $\theta_{k_{max}}$ the curvature decreases. This may appear to be a drawback for using this type of segment in corner smoothing, however, in this application, the polar angle is restricted to the range of $0<\theta<\theta_{k_{max}}$ so the curvature is monotonically increasing along the transition segment. 

The last curve property of interest is the course angle which is given by
\begin{align}
\psi\left(\theta\right) & =\theta+\arctan\left(2\theta\right).\label{eq:fermat_course_angle}
\end{align} 
For the corner smoothing application, it is important to define equations for a reflected segment that starts at an arbitrary point along the segment and tracks back to the start of the segment. This operation is more involved for a Fermat Spiral segment than the other segments defined in this section so it will receive special attention here. The position equation for the reflected Fermat Spiral is parameterized by the curve end point $\left(x_{end}, y_{end}\right)$, and course at the end of the segment $\psi_{end}$ and is given by
\begin{eqnarray}
x\left(\theta\right)=&x_{end}+c\sqrt{\theta_{end}-\theta}\cos\left(\rho\left(\theta-\theta_{end}\right)+\psi_{end}\right) \label{eq:fermat_positionx_ref} \\
y\left(\theta\right)=&y_{end}+c\sqrt{\theta_{end}-\theta}\sin\left(\rho\left(\theta-\theta_{end}\right)+\psi_{end}\right) \label{eq:fermat_positiony_ref}
\end{eqnarray}
and $\theta_{end}$ is the polar angle at the end of the reflected Fermat Spiral segment.

Lekkas et al. \cite{lekkas_continuous-curvature_2013}, show how the shaping parameter, $c$, can be set such that the maximum curvature of the segment is equal to the maximum curvature of the aircraft. This is accomplished by solving \eqref{eq:fermat_curvature} and using $\theta = \theta_{k_{max}}$ and $k\left(\theta_{k_{max}}\right)=k_{max}$. Then an expression for $c$ is given by

\begin{align}
c & =\frac{1}{k_{max}}\frac{2\sqrt{\theta_{k_{max}}}\left(4\theta_{k_{max}}^{2}+3\right)}{\left(4\theta_{k_{max}}^{2}+1\right)^{\frac{3}{2}}}\label{eq:fermat_c}
\end{align}
which will result in a scaled curve that reaches a maximum curvature of $k_{max}$. Thus, the Fermat spiral can be parameterized to respect the same maximum curvature constraint used in the clothoid smoothing implementation. 

One drawback to using the Fermat spiral for corner smoothing is that the length of the path is given by
\begin{align}
s & =c\sqrt{2}\int_{0}^{\theta_{k_{max}}}\sqrt{1+4\theta^{2}}d\theta\label{eq:fermat_length}
\end{align}
which does not have a closed form solution. In \cite{lekkas_continuous-curvature_2013} the length of a Fermat segment is expressed as a hyper-geometric function with guaranteed convergence over the domain of interest. Thus, in implementation, the computational cost of \eqref{eq:fermat_length} can be mitigated using a lookup table or an iterative solution to the hyper-geometric function with some convergence criteria.

%%%%%%%%
% Curvilinear Motion Theory
%%%%%%%%
\subsection{Curvilinear Motion Theory\label{sec:background_curv}}
Curvilinear motion theory defines the motion of a particle along a curved path \cite{potter_dynamics}. Properties of this theory will be used to relate the geometry of the path with the aircraft state vector. The velocity and acceleration along a curved path is given by 
\begin{equation}
\boldsymbol{v}=\dot{s}\boldsymbol{e}_{t}\label{eq:curvlinear_velocity}
\end{equation}
and 
\begin{eqnarray}
	\boldsymbol{a} & = & \ddot{s}\boldsymbol{e}_{t}+\dot{s}\dot{\psi}\boldsymbol{e}_{n}
\end{eqnarray}
% \begin{eqnarray}
% \boldsymbol{a} & = & \ddot{s}\boldsymbol{e}_{t}+\dot{s}\frac{d\boldsymbol{e}_{t}}{dt} \\
% & = & \ddot{s}\boldsymbol{e}_{t}+\ddot{s}k\boldsymbol{e}_{n} \\
% & = & \ddot{s}\boldsymbol{e}_{t}+\dot{s}\dot{\phi}\boldsymbol{e}_{t} \\
% \end{eqnarray}
where $\boldsymbol{e}_{t}$ and $\boldsymbol{e}_{n}$, are the tangent and normal vectors, and $\dot{\psi}$ is the angular rate of the path. 
\subsection{Reverse INS Integration \label{subsec:RII}}
A primary objective of this work is to generate body frame specific force and angular rates experienced by the aircraft. These quantities are used to represent measurements from an IMU. This section describes a method for generating the specific force and angular rates from aircraft states called Reverse INS Integration (RII). The RII method solves the INS kinematic equations for the body frame accelerations and angular rates.

The objective is to provide expressions for the specific force, $\boldsymbol{f^b}$, and angular rate, $\boldsymbol{\omega_{ib}^b}$, vectors that represent measurents from an IMU. Farrell \cite{farrell_aided_2008} provides expressions for these quantities. After applying a non-rotating earth assumption and solving for the quantities of interest they are given by
\begin{eqnarray}
	[\boldsymbol{\omega_{ib}^{b}}(t_k)\times] & = & \frac{\ln\left[R_{b}^{n}(t_{k-1})^{T}R_{b}^{n}(t_{k})\right]}{dt} \label{eq:Omega_ib_final} \\
	\boldsymbol{f^{b}}(t_k) & = & R_{n}^{b}(t_k)\left[\boldsymbol{\dot{v}_{e}^{n}}(t_k)-\boldsymbol{g^{n}}(t_k)\right]\label{eq:spec_force}
\end{eqnarray}
where $R_n^b$ is the rotation matrix from the navigation frame to the body frame, $\boldsymbol{g^n}$ is the gravity vector in the navigation frame. For the RII method, the time derivative of the velocity ($\boldsymbol{\dot{v}_e^n}$) is computed using a first order approximation that is susceptible to error for large timesteps.

%% file: figures/path_smoothing.tex
\tikzset{every picture/.style={line width=0.75pt}} %set default line width to 0.75pt        

\begin{tikzpicture}[x=0.75pt,y=0.75pt,yscale=-1,xscale=1]
%uncomment if require: \path (0,340); %set diagram left start at 0, and has height of 340

%Straight Lines [id:da6341609088022722] 
\draw [color={rgb, 255:red, 0; green, 0; blue, 0 }  ,draw opacity=1 ]   (59.44,254.34) -- (147.5,254.34) ;
\draw [shift={(149.5,254.34)}, rotate = 180] [color={rgb, 255:red, 0; green, 0; blue, 0 }  ,draw opacity=1 ][line width=0.75]    (10.93,-4.9) .. controls (6.95,-2.3) and (3.31,-0.67) .. (0,0) .. controls (3.31,0.67) and (6.95,2.3) .. (10.93,4.9)   ;
%Straight Lines [id:da9087576499901888] 
\draw [color={rgb, 255:red, 0; green, 0; blue, 0 }  ,draw opacity=1 ]   (59.44,254.34) -- (59.44,160.86) ;
\draw [shift={(59.44,158.86)}, rotate = 450] [color={rgb, 255:red, 0; green, 0; blue, 0 }  ,draw opacity=1 ][line width=0.75]    (10.93,-4.9) .. controls (6.95,-2.3) and (3.31,-0.67) .. (0,0) .. controls (3.31,0.67) and (6.95,2.3) .. (10.93,4.9)   ;
%Straight Lines [id:da41518025270878334] 
\draw  [dash pattern={on 4.5pt off 4.5pt}]  (98.3,239.72) -- (221.37,224.03) -- (239.2,221.75) ;
%Straight Lines [id:da887150084624639] 
\draw  [dash pattern={on 4.5pt off 4.5pt}]  (134.65,148.14) -- (239.2,221.75) ;
%Straight Lines [id:da0748360420096934] 
\draw [color={rgb, 255:red, 144; green, 19; blue, 254 }  ,draw opacity=1 ]   (98.3,239.72) -- (188.71,228.21) ;
%Straight Lines [id:da5898329897510202] 
\draw [color={rgb, 255:red, 144; green, 19; blue, 254 }  ,draw opacity=1 ]   (134.65,148.14) -- (200.89,194.91) ;
%Curve Lines [id:da9172611722494899] 
\draw [color={rgb, 255:red, 208; green, 2; blue, 27 }  ,draw opacity=1 ]   (188.71,228.21) .. controls (211.26,225.65) and (224.33,210.84) .. (200.89,194.91) ;
%Flowchart: Connector [id:dp939708556361839] 
\draw  [fill={rgb, 255:red, 0; green, 0; blue, 0 }  ,fill opacity=1 ] (95.59,239.21) .. controls (95.97,237.98) and (97.49,237.21) .. (98.99,237.49) .. controls (100.49,237.77) and (101.39,239) .. (101.01,240.23) .. controls (100.63,241.46) and (99.1,242.22) .. (97.61,241.94) .. controls (96.11,241.66) and (95.21,240.43) .. (95.59,239.21) -- cycle ;
%Flowchart: Connector [id:dp5863733768614277] 
\draw  [fill={rgb, 255:red, 0; green, 0; blue, 0 }  ,fill opacity=1 ] (131.94,147.63) .. controls (132.33,146.4) and (133.85,145.63) .. (135.35,145.92) .. controls (136.84,146.2) and (137.75,147.42) .. (137.36,148.65) .. controls (136.98,149.88) and (135.46,150.65) .. (133.96,150.37) .. controls (132.46,150.08) and (131.56,148.86) .. (131.94,147.63) -- cycle ;
%Flowchart: Connector [id:dp16398918856225886] 
\draw  [fill={rgb, 255:red, 0; green, 0; blue, 0 }  ,fill opacity=1 ] (236.49,221.24) .. controls (236.87,220.01) and (238.4,219.25) .. (239.89,219.53) .. controls (241.39,219.81) and (242.3,221.03) .. (241.91,222.26) .. controls (241.53,223.49) and (240.01,224.26) .. (238.51,223.98) .. controls (237.01,223.7) and (236.11,222.47) .. (236.49,221.24) -- cycle ;
%Straight Lines [id:da7301130618012559] 
\draw  [dash pattern={on 0.84pt off 2.51pt}]  (239.2,221.75) -- (271,218) ;
%Shape: Arc [id:dp7097534510225836] 
\draw  [draw opacity=0] (223.73,208.86) .. controls (227.43,204.44) and (232.99,201.63) .. (239.2,201.63) .. controls (249.31,201.63) and (257.68,209.06) .. (259.13,218.75) -- (239.2,221.75) -- cycle ; \draw   (223.73,208.86) .. controls (227.43,204.44) and (232.99,201.63) .. (239.2,201.63) .. controls (249.31,201.63) and (257.68,209.06) .. (259.13,218.75) ;

% Text Node
\draw (168.43,267.93) node  [font=\small,color={rgb, 255:red, 0; green, 0; blue, 0 }  ,opacity=1 ]  {$x$};
% Text Node
\draw (44.24,170.79) node  [font=\small,color={rgb, 255:red, 0; green, 0; blue, 0 }  ,opacity=1 ]  {$y$};
% Text Node
\draw (161.83,235.27) node [anchor=north west][inner sep=0.75pt]    {$\boldsymbol{p_{1}}$};
% Text Node
\draw (173,149.4) node [anchor=north west][inner sep=0.75pt]    {$\boldsymbol{p_{2}}$};
% Text Node
\draw (242,179.4) node [anchor=north west][inner sep=0.75pt]    {$\psi _{\Delta }$};

\end{tikzpicture}

%% file: sections/PathSmoothing.tex
In the first stage of the aircraft state generator presented in this work, a piece-wise linear path is converted into a smooth path with continuous course angle or curvature. In the case of continuous curvature, a limit on curvature and curvature rate can be imposed to ensure that the path will be flyable by a specific aircraft. In this way, the physical limitations of the aircraft can be considered and the resulting states can more accurately reflect the behavior of the aircraft.

This section describes three methods for smoothing a piece-wise linear path 1) Arc Fillet, 2) Clothoid Fillet, and 3) Fermat Fillet. Each of these methods perform corner smoothing at every way-point in the path. The resulting path will consist of a series of straight line, arc, and transition segments with varying curvature. This section is particularly useful for a person in the navigation field that may be unfamiliar with corner smoothing algorithms because it provides a unified framework for three methods used in the path planning field.

The path smoothing algorithms presented in this section assume that the spacing between the way-points is sufficiently large enough to perform the indicated corner smoothing method. For example, if the start or end way-point in Fig. \ref{fig:smoothing_example} was closer to the corner point than the connection between the straight path and the smoothing arc, then the algorithms presented here would not be sufficient.

%\subsection{Path Smoothing Algorithm\label{subsec:Smoothing-Algorithm}}
The first path smoohing approach presented in this section is the Arc
Fillet method where a maximum curvature segment is used to connect
two straight segments. This method has continuous course angle but
has discontinuous path curvature. The second is the Clothoid Fillet
method where clothoid segments are used to smooth the corners or, 
where needed, transition to a $k_{max}$ segment. This method has
continuous curvature and course angle and respects limits on curvature
and curvature rate. The third is the Fermat Fillet method which is
similar to the Clothoid Fillet method but Fermat spiral segments are
used in place of clothoid segments. This method has continuous curvature
and course angle and respects limits on curvature, however, it is
not guaranteed to respect limits on curvature rate. These properties
are summarized in Table \ref{tab:method_comparison}.

\begin{table}
\centering{}\caption{Path smoothing comparison\label{tab:method_comparison}}
\def\arraystretch{1.25} % Add vertical buffer to table for readability
\begin{tabular}{c|c|c|c|c}
 & \multicolumn{2}{c|}{Continuous} & \multicolumn{2}{c}{Respect}\tabularnewline
Method & $\psi$ & $k$ & $k_{max}$ & $k'_{max}$\tabularnewline
\hline 
Arc Fillet      & X & - & X & - \tabularnewline
Clothoid Fillet & X & X & X & X\tabularnewline
Fermat Fillet   & X & X & X & - \tabularnewline
\end{tabular}
\end{table}

The three approaches presented in this section iterate through the corners in the path and compute a series of path segments that represent the resulting smooth path. The sharpness of the corner and positions of the way-points before and after the corner are required to determine the smoothing segments. The position of the way-points is provided to the algorithm. The sharpness of the corner is represented by the change of course angle, $\psi_\Delta$, required to transition from the entry segment to the exit segment. The $\psi_\Delta$ quantity can be computed by expressing the entry and exit segments as vectors ($\boldsymbol{p_1}$ and $\boldsymbol{p_2}$ from Fig. \ref{fig:smoothing_example}) and using properties of the dot product to determine the angle between the vectors. Another important quantity for each corner is the turn direction, $\rho$, which is determined using the sign of the cross product of the two vectors. The sign convention for $\rho$ is given in \eqref{eq:direction}.

\subsection{Arc Fillet\label{subsec:Arc-Fillet}}

\begin{figure}
\noindent\begin{minipage}[t]{1\columnwidth}%
\begin{center}
\input{figures/arc_fillet.tex}
\par\end{center}%
\end{minipage}
\caption{Arc Fillet example where the origin for the plot is at the connection point along the straight segment entering the way-point and the initial course angle for the arc matches the course angle of the entering segment. \label{fig:arc_fillet}}
\end{figure}
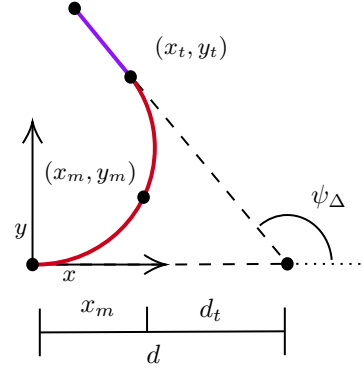

The Arc Fillet method smooths a corner at a way-point by adding a circular arc of maximum curvature ($k_{max}$) that transitions from the first straight segment to the other. The arc segment is parameterized by the radius, $r=1/k_{max}$, and the arc length, $s=r\psi_\Delta$, as discussed in Section \ref{subsec:Arc}.
Then \eqref{eq:arc_eqs} is used with $\theta = \psi_\Delta$ to determine the end point, $\left(x_{t},y_{t}\right)$, and the midpoint, $\left(x_{m},y_{m}\right)$, of the arc. The distance along the x-axis from the midpoint of the arc is computed geometrically as
\begin{eqnarray}
d_{t} & = & \text{\ensuremath{\left|\frac{y_{m}}{\tan^{-1}\left(\frac{\pi-\psi_{\Delta}}{2}\right)}\right|}}\label{eq:dt_connection}
\end{eqnarray}
and the distance from the way-point to the connection point of the
arc on the first straight segment is given by
\begin{eqnarray*}
d & = & d_{t}+x_{m}.
\end{eqnarray*}
This distance is used to determine the connection point for the arc segment on the first straight segment. Fig. \ref{fig:arc_fillet} shows the geometry of the Arc Fillet method and includes a graphical representation of $x_{m}$ and $d_{t}$.

 \begin{comment}
The equations above are used to define a circular arc starting at the origin to smooth the corner of a waypoint. This can be used as a base-case then the resulting arc can be translated and rotated to smooth a way-point\textbf{CAN WE DESCRIBE THIS MORE CLEARLY OR CAN WE SKIP DISCUSSING THE TRANSLATION AND ROTATION}  Given a translation vector, $\boldsymbol{p_{t}}$, and a rotation angle $\psi_{r}$, the rotation matrix is defined as
\begin{equation}
R=\left[\begin{array}{cc}
\cos(\psi_{r})\quad & -\sin(\psi_{r})\\
\sin(\psi_{r})\quad & \cos(\psi_{r})
\end{array}\right]\label{eq:rot_matrix}
\end{equation}
and the initial parameters of the arc are changed according to \textbf{DROGE SAID THIS LOOKS WEIRD, WHY DO WE TRANSLATE FIRST?? AND I DON'T LIKE THE 0 SUBSCRIPTS EITHER...}
git 
\begin{equation}
\boldsymbol{p_{0_{n}}}=R(\boldsymbol{p_{0}}+\boldsymbol{p_{t}})\label{eq:translate_and_rotate_1}
\end{equation}

\begin{equation}
\psi_{0_{n}}=\psi_{0}+\psi_{r}\label{eq:translate_and_rotate_2}
\end{equation}
With the initial point and course angle changed for the arc segment,
the segment is effectively rotated and translated into place along
the path. Then \eqref{eq:arc_eqs} is used to sample the path.
\end{comment}

\subsection{Clothoid Fillet\label{subsec:Clothoid-Fillet}}
The Clothoid Fillet method uses clothoid segments to connect the straight segments and arcs while maintaining continuous curvature. Using clothoid segments to generate continuous curvature paths is well documented in \cite{fraichard_reeds_2004}. This section will describe the specific application of smoothing corners using a clothoid fillet method within the unified framework of the current work. 

\begin{figure}
\noindent\begin{minipage}[t]{0.85\columnwidth}%
\begin{center}
\input{figures/clothoid_segments.tex}
\par\end{center}%
\end{minipage}
\caption{Smoothing with clothoid segments. The blue lines represent clothoids
reflected about the bisecting angle and the red segment is a maximum
curvature segment. \label{fig:smoothing_with_clothoids}}
\end{figure}
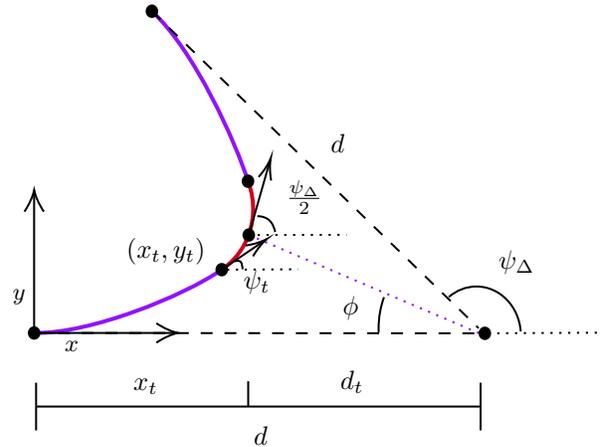

The Clothoid Fillet method relies on the fact that the scaling factor, $\sigma_c$, of a clothoid segment represents the rate of change of the curvature of the segment. This method uses the maximum change in curvature, $k'_{max}$, to transition from the initial straight segment ($k=0$), to a maximum curvature ($k=k_{max}$) arc. The transition segment is then reflected to obtain a transition from the constant curvature arc to the second straight segment. The resulting smoothed path consists of a series of segments including straight segments, $k_{max}$ orbits, and clothoid transition segments with curvature changing linearly by $k'_{max}$. Fig. \ref{fig:smoothing_with_clothoids} shows a representation of a series of segments generated by the path smoothing algorithm when transition segments are used. The blue segments are transition segments and the red segment is a $k_{max}$ arc.

When the turn is small enough enough two clothoid segments can be used to smooth the corner, otherwise, a $k_{max}$ arc must be added. The course angle change at which the $k_{max}$ arc segment is required is determined by solving \eqref{eq:clothoid_curvature} with $k_0=0$ to determine the segment length at which the segment reaches maximum curvature ($k=k_{max}$) as
\begin{equation}
    s_{kmax} = \frac{k_{max}}{k_{max}^\prime}. \label{eq:skmax}
\end{equation}
Then evaluating \eqref{eq:clothoid_heading} with $k_0=\psi_0=0$ and $s=s_{kmax}$ gives
\begin{equation}
\psi_{m}  =  \frac{k_{max}^{2}}{2k_{max}^{\prime}}\\
\end{equation}
where $\psi_{m}$ is the change of course angle produced change by one clothoid segment. Thus, the maximum course angle change that can be spanned with two clothoid segments is $2\psi_{m}$. 

\begin{comment}
Combining \eqref{eq:clothoid_heading} and \eqref{eq:clothoid_curvature} and solving for the course angle associated with  
To determine the maximum change in course angle that can be smoothed with two clothoid segments, combine equations 

So the first step in the clothoid fillet method is to determine when a $k_{max}$ arc must added. This is determined by starting with \eqref{eq:clothoid_curvature} with $k_{0}=0$. Then substituting the path length, $s$, for the $k_{max}$ condition results in $s_{kmax}=k_{max}/k_{max}^{\prime}$. Then consider \eqref{eq:clothoid_heading} with $k_{0}=\psi_{0}=0$ and using the formula for $s_{kmax}$ yields \textbf{CONFUSING??}
\end{comment}
\begin{comment}
\begin{eqnarray*}
\psi_{m} & = & \frac{1}{2}k_{max}^{\prime}s_{kmax}^{2}\\
 & = & \frac{1}{2}k_{max}^{\prime}\frac{k_{max}^{2}}{k_{max}^{\prime2}}\\
 & = & \frac{k_{max}^{2}}{2k_{max}^{\prime}}
\end{eqnarray*}
\end{comment}

It remains to define the parameters of the line segments used for corner smoothing with the Clothoid Fillet method. The parameters include initial position, $(x_0,y_0)$, initial course angle, $\psi_0$, initial curvature, $k_0$, curvature rate, $\sigma_c$, and segment length, $s$. The following paragraphs discuss how each of these parameters are determined for the smoothing segments based on the course angle change of the corner.

The initial position for the first clothoid segment is the attachment point for the smoothing segments to the initial straight segment. The attachment point can only be determined after the smoothing segments are defined so the smoothing segments will be defined for a base-case where the initial position, course angle, and curvature for the first clothoid segment are $x_0=y_0=\psi_0=k_0=0$. 

The clothoid curvature rate, $\sigma_c$, for the initial clothoid segment is set to the maximum curvature rate ($\sigma_c=k^\prime_{max}$) desired for the current application. The second clothoid segment is a reflection of the first clothoid segment so the curvature rate is negated ($\sigma_c=-k^\prime_{max}$) to reduce the curvature from the transition curvature back to $k=0$.

\begin{figure}
\noindent\begin{minipage}[t]{1\columnwidth}%
\begin{center}
\input{figures/arc_calc.tex}
\par\end{center}%
\end{minipage}
\caption{Definition of key parameters in the development of a $k_{max}$ segment.
\label{fig:arc_length_calc}}
\end{figure}
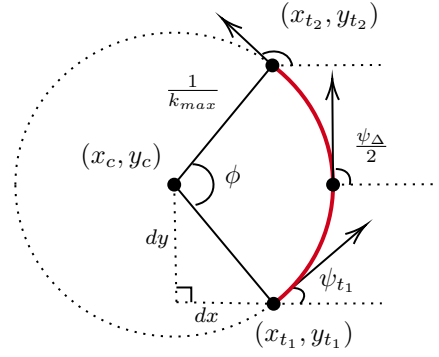

The segment length for each of the smoothing segments is determined using the course angle change of the corner and the curvature rate for the clothoid segments. For the case where the turn can be spanned by two clothoid segments ($\psi_{\Delta}\leq2\psi_{m}$), \eqref{eq:clothoid_heading} is solved for $s$ with $\psi=\frac{\psi_{\Delta}}{2}$, $\psi_{0}=0$, and $\sigma_c=k'_{max}$ which yields $s=\sqrt{\psi_{\Delta}/k_{max}^{\prime}}$. For the case where a maximum curvature segment is required to smooth the corner, $s=s_{kmax}$ (see \eqref{eq:skmax}). The segment length defined here is the same for both clothoids segments used in the Clothoid Fillet Method. The segment length of the arc segment when one is required is computed using the arc length formula as
\begin{equation}
    s = \frac{\phi}{k_{max}} \label{eq:arc_s}
\end{equation}
where $\phi$ is the central angle of the arc (see Fig. \ref{fig:arc_length_calc}) and is expressed as
\begin{equation}
    \phi = 2\left(\frac{\psi_\Delta}{2}-\psi_t\right).
\end{equation}

For small turns ($\psi_\Delta \leq 2\psi_m$), two clothoid segments are defined to smooth the corner. Each clothoid segment provides a course angle change of $\psi_\Delta/2$. The initial position ($x_t$,$y_t$) and course angle ($\psi_t$) of the second clothoid segment is determined by evaluating \eqref{eq:fresnel_x}-\eqref{eq:clothoid_curvature} with $\sigma_c=k^\prime_{max}$, $s=\sqrt{\psi_\Delta/k^\prime_{max}}$, and $k_0=0$. The initial curvature of the second clothoid segment is given by $k_0=\sigma_c s=\sqrt{k^\prime_{max}\psi_\Delta}$.

For large turns ($\psi_\Delta > 2\psi_m$), a $k_{max}$ arc is added between the clothoid transition segments. In this case, the initial clothoid segment transitions from zero curvature to $k_{max}$ so the length of the segment is $s_{kmax}$ (see \eqref{eq:skmax}). The initial position (${x_t}_1$,${y_t}_1$) and course angle (${\psi_t}_1$) of the arc segment is then determined by evaluating \eqref{eq:fresnel_x}-\eqref{eq:clothoid_curvature} with $c=k^\prime_{max}$, $s=s_{kmax}$, and $k_0=0$. The length of the arc segment is given in \eqref{eq:arc_s}. The initial position (${x_t}_2$,${y_t}_2$) and course angle (${\psi_t}_2$) of the second clothoid segment can be determined by evaluating \eqref{eq:arc_x}-\eqref{eq:arc_k} with $(x_0,y_0,\psi_0)=({x_t}_1,{y_t}_1,{\psi_t}_1)$, $\theta = \phi$, and $r=1/k_{max}$. 

The parameters defined in the preceding paragraphs are summarized in Table \ref{tab:segment_definitions}. The table shows the segment parameters required to smooth large and small corners using the Clothoid Fillet method. The parameters are provided in terms of the course angle change of the corner ($\psi_\Delta$), the maximum curvature of the smoothed corner ($k_{max}$) and the maximum curvature rate ($k^\prime_{max}$).

\begin{table}
\caption{Segment Definition for Clothoid Fillet Corner Smoothing. \label{tab:segment_definitions}}
\centering{}%
\def\arraystretch{1.75} % Add vertical buffer to table for readability
\begin{tabular}{m{0.4cm}|m{1cm}|m{1.5cm}|m{1.25cm}|m{0.75cm}|m{1cm}}
                             $\boldsymbol{\psi_\Delta}$ & \textbf{Segment}  & ($\boldsymbol{x_0}$,$\boldsymbol{y_0}$,$\boldsymbol{\psi_0}$) & $\boldsymbol{k_0}$  & $\boldsymbol{\sigma_c}$              & $\boldsymbol{s}$            \\ \hline
\multirow{2}{*}{\rotatebox[origin=c]{90}{\textbf{Small}}} & clothoid & ($x_a$,$y_a$,$\psi_a$) & 0       & $k^\prime_{max}$ & $\sqrt{\frac{\psi_{\Delta}}{k_{max}^{\prime}}}$          \\
                            & clothoid & ($x_t$,$y_t$,$\psi_t$) & $\sqrt{k^\prime_{max}\psi_\Delta}$ & $-k^\prime_{max}$               & $\sqrt{\frac{\psi_{\Delta}}{k_{max}^{\prime}}}$          \\
                            \hline
\multirow{3}{*}{\rotatebox[origin=c]{90}{\textbf{Large}}} & clothoid & ($x_a$,$y_a$,$\psi_a$) & 0       & $k^\prime_{max}$ & $\frac{k_{max}}{k_{max}^\prime}$          \\
                            & arc      & ($x_t$,$y_t$,$\psi_t$) & $k_{max}$ & 0 & $\frac{\phi}{k_{max}}$ \\
                            & clothoid & ($x_{t_2}$,$y_{t_2}$,$\psi_{t_2}$) & $k_{max}$ & $-k^\prime_{max}$                & $\frac{k_{max}}{k_{max}^\prime}$     
\end{tabular}
\begin{comment}
\begin{tabular}{c|c|c|c|c|c}
\textbf{Segment} & $\boldsymbol{p_{0}}$ & $\psi_{0}$ & $k_{0}$ & $k^{\prime}$ & $s$\tabularnewline
\hline 
Initial segment & $\boldsymbol{0}$ & 0 & 0 & $k_{max}^{\prime}$ & $s_{c}$\tabularnewline
$k_{max}$ segment & $\boldsymbol{p_{t}}$ & $\psi_{t}$ & $k_{max}$ & 0 & $\frac{\phi}{k_{max}}$\tabularnewline
Final segment & $\boldsymbol{p_{t}}$ & $\psi_{t}$ & $s_{c}k_{max}^{\prime}$ & $-k_{max}^{\prime}$ & $s_{c}$\tabularnewline
\end{tabular}
\end{comment}
\end{table}

With the smoothing segments defined for the base case of the Clothoid Fillet method, it remains to determine the attachment point ($x_a,y_a,\psi_a$) on the initial straight segment. The attachment point is determined by computing the parameter $d$ in Fig. \ref{fig:smoothing_with_clothoids} which represents the distance from the attachment point to the end of the straight segment. Equation \eqref{eq:dt_connection} can be used to compute the value for $d_{t}$ where $y_{m}$ is the $y$ component of the transition point between the two smoothing clothoid segments ($y_m=y_t$), or the midpoint of the $k_{max}$ segment, where one is used. Then the attachment distance \textbf{$d$} is calculated as
\begin{equation}
d=x_{t}+d_{t}
\end{equation}
where $x_t$ is the $x$ component of the transition point between the two clothoid segments, or the midpoint of the $k_{max}$ arc segment, where one is used. Then \eqref{eq:line_x} and \eqref{eq:line_y} can be evaluated using the segment parameters of the initial straight segment and with $s=s_o-d$ where $s_o$ is the original length of the initial straight segment.

Finally, the initial position and heading for each of the smoothing segments are adjusted based on the calculated attachment position and course angle. The adjustment is accomplished by adding the attachment position and course angle to the initial position and course angle of each of the segments.

\subsection{Fermat Fillet}
The Fermat spiral was shown to be an effective transition segment for path smoothing applications in \cite{lekkas_continuous-curvature_2013}. This section reiterates some of the development from \cite{lekkas_continuous-curvature_2013} to fit in the framework presented in this work and define the segments required to smooth a corner using the Fermat Fillet method. The parameters required to define a Fermat spiral segment are the initial position and course angle ($x_0, y_0, \psi_0$), initial curvature ($k_0$), and span of the polar angle ($\theta_\Delta$).

The Fermat Fillet method results in a continuous curvature path where the transition segments are Fermat spirals. Similar to the Clothoid Fillet method, the corner can be smoothed with two Fermat spirals for small turns, or two Fermat spirals and a $k_{max}$ arc segment for large turns. The largest course angle change that can be achieved with one Fermat spiral segment is determined by combining \eqref{eq:theta_kmax} and \eqref{eq:fermat_course_angle} as 
\nolinebreak
\begin{eqnarray*}
\psi_{m} & = & \sqrt{\frac{\sqrt{7}}{2}-\frac{5}{4}}+\tan^{-1}\left(\sqrt{\frac{\sqrt{7}}{2}-\frac{5}{4}}\right)\\
 & \approx & 0.7650\quad\text{rad}.
\end{eqnarray*}
\begin{comment}
The transition segments can span a corner with two segments (for a small enough $\psi_{\Delta}$) or transition to a $k_{max}$ arc segment (for larger $\psi_{\Delta}$). It remains to define the conditions on which a $k_{max}$ segment is required, and the equations used to compute the transition parameters (i.e. $(x_{t},y_{t})$, $\psi_{t}$). 

Using \eqref{eq:theta_kmax} and \eqref{eq:fermat_course_angle}
the course angle change when the Fermat spiral reaches $k_{max}$
can be computed as
\begin{eqnarray*}
\psi_{m} & = & \sqrt{\frac{\sqrt{7}}{2}-\frac{5}{4}}+\tan^{-1}\left(\sqrt{\frac{\sqrt{7}}{2}-\frac{5}{4}}\right)\\
 & \approx & 0.7650\quad\text{rad}
\end{eqnarray*}
\end{comment}
Then the largest course angle change ($\psi_{\Delta}$) that can be spanned by two Fermat
spiral segments is $2\psi_{m}$. 

In the case where a $k_{max}$ segment is needed ($\psi_{\Delta}>2\psi_{m}$), $\theta_{kmax}$ is defined by  \eqref{eq:theta_kmax} and the course angle at the transition between the first Fermat spiral segment and the $k_{max}$ arc is $\psi_{m}$. In the case where two Fermat spiral segments can span $\psi_{\Delta}$ (where $\psi_{\Delta}\leq2\psi_{m}$), the course angle at the transition is given by $\psi_{t}=\psi_{\Delta}/2$. In this case, $\theta_{kmax}$ is more difficult to compute because  \eqref{eq:fermat_course_angle} is not invertible. However, \cite{lekkas_continuous-curvature_2013} suggests using a root finding method to iteratively solve for the roots of the course angle equation. This can be implemented by defining
\begin{eqnarray*}
f(\theta) & = & \theta+\tan^{-1}(2\theta)-\psi_{t}
\end{eqnarray*}
where subtracting $\psi_{t}$ moves the root of the course angle equation
such that applying the root finding method will result in the $\theta$ associated with the desired course angle $\psi_{t}$. 
\begin{comment}
In practice, applying Halley's method requires iteratively solving
\begin{eqnarray}
\theta_{n+1} & = & \theta_{n}-\frac{2f(\theta_{n})f'(\theta_{n})}{2[f'(\theta_{n})]^{2}-f(\theta_{n})f''(\theta_{n})} \label{eq:halleys}
\end{eqnarray}
until some convergence criteria is met ($|\theta_{n+1}-\theta_{n}|<\epsilon$),
then $\theta_{kmax}=\theta_{n+1}.$ In this work the convergence value
was chosen as $\epsilon=1e^{-6}$, and given the range of $\psi_{t}$
$\left(0<\psi_{t}\leq\psi_{m}\right)$. Halley's method is guaranteed
to converge, and converges in just a few iterations. 
\end{comment}

With $\theta_{kmax}$ and $\psi_{t}$ defined, the first Fermat spiral
segment can be completely defined. The base case starts at the origin
with $\psi_{0}=0$, and $\theta_{max}=\theta_{kmax}$ using the value
for $\theta_{kmax}$ determined above. Then, the scale factor, $c$,
in the Fermat spiral can be computed using \eqref{eq:fermat_c}
with $\theta_{kmax}=\psi_{m}$ which ensures that $k_{max}$ will
be achieved at $\psi_{m}$ even if the segment ends prior to the curvature reaching
$k_{max}$. The transition point, $\text{(\ensuremath{x_{t}},\ensuremath{y_{t}})}$,
can be determined by evaluating \eqref{eq:fermat_positionx} and \eqref{eq:fermat_positiony}
at $\theta_{max}$ and the course angle at the transition is $\psi_{t}$.

In the case where a $k_{max}$ segment is required, an arc is added
with with an initial position of $(x_{t},y_{t})$, initial course
angle of $\psi_{t}$, radius of $1/k_{max}$, and arc length given
by
\begin{eqnarray*}
s & = & \frac{\psi_{\Delta}-2\psi_{t}}{k_{max}}
\end{eqnarray*}
then \eqref{eq:arc_eqs} can be evaluated at $\theta=\psi_{\Delta}-2\psi_{t}$
to get the transition point for the second Fermat spiral segment,
$(x_{t},y_{t})$, and $\theta=\frac{\psi_{\Delta}-2\psi_{t}}{2}$
to get the midpoint of the arc, $(x_{m},y_{m})$.

\begin{figure}
\noindent\begin{minipage}[t]{0.85\columnwidth}%
\begin{center}
\input{figures/fermat_fillet.tex}
\par\end{center}%
\end{minipage}
\caption{Parameters used in Fermat Fillet method. \label{fig:fermat_fillet}}
\end{figure}
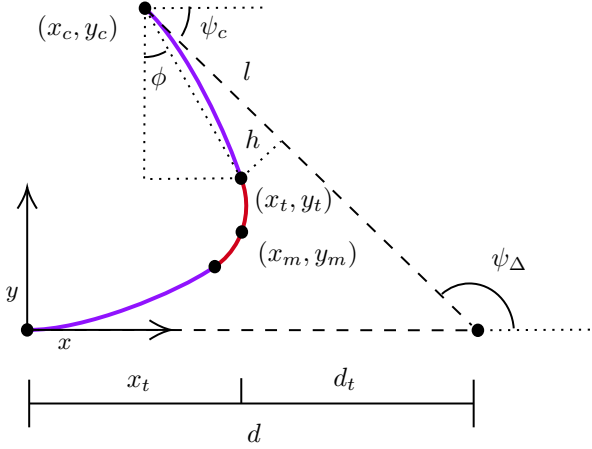

The second transition segment is then defined that connects to the
first transition segment, or the $k_{max}$ arc at $(x_{t},y_{t})$.
The second transition segment is a reflection of the first transition
segment. For the Fermat spiral, this reflection is accomplished by
computing the connection point, $(x_{c},y_{c})$, to the exit segment
and flipping the sign of the direction parameter, $\rho$. The connection point is determined by computing the deviation from the straight line segment, where the cross track deviation, $h$, and down track deviation, $l$, are given by
\nolinebreak
\begin{eqnarray*}
h & = & c\sqrt{\theta_{kmax}}\sin\text{\ensuremath{\left(\ensuremath{\theta_{kmax}}\right)}}\\
l & = & c\sqrt{\theta_{kmax}}\cos\text{\ensuremath{\left(\theta_{kmax}\right)}}
\end{eqnarray*}
\nolinebreak
then the magnitude of the deviation is given by
\nolinebreak
\begin{eqnarray*}
\beta & = & \sqrt{l^{2}+h^{2}}.
\end{eqnarray*}
\nolinebreak
Then determine the angle, $\phi$, that relates $\left(x_{c},y_{c}\right)$ and $\left(x_{t},y_{t}\right)$ as follows
\nolinebreak
\begin{eqnarray*}
\psi_{c} & = & \pi-\psi_{\Delta}\\
\phi & = & \frac{\pi}{2}-\psi_{c}-\theta_{kmax}.
\end{eqnarray*}
The connection point is then calculated as
\begin{eqnarray*}
x_{c} & = & x_{t}-\beta\sin\left(\phi\right) \\
y_{c} & = & y_{t}+\rho\beta\cos\left(\phi\right).
\end{eqnarray*}
The geometry used to compute these parameters is shown graphically in Fig. \ref{fig:fermat_fillet}. These results are used with \eqref{eq:fermat_positionx_ref} and \eqref{eq:fermat_positiony_ref} to compute the properties of the reflected segment.

% Table \ref{tab:fermat_fillet_segment_definitions} provides a summary of the parameters that define the segments involved in the Fermat Fillet method.
This method calculates a final connection point and uses a reflection parameter so that the terminal point of the final segment is at the transition point, $(x_t, y_t)$. This requires additional computation for the creation and sampling of the segments but provides simple segment parameter definitions. 

\begin{table}
	\caption{Segment Definition for Fermat Fillet Corner Smoothing. \label{tab:fermat_fillet_segment_definitions}}
	\centering{}%
	\def\arraystretch{1.75} % Add vertical buffer to table for readability
	\begin{tabular}{m{0.4cm}|m{1cm}|m{1.5cm}|m{0.5cm}|m{1cm}|m{1cm}|m{0.25cm}}
		$\boldsymbol{\psi_\Delta}$ & \textbf{Segment}  & ($\boldsymbol{x_0}$,$\boldsymbol{y_0}$,$\boldsymbol{\psi_0}$) & $\boldsymbol{k_0}$ &$\boldsymbol{s}$ & $\boldsymbol{\theta_{kmax}}$ \textbf{Method} & \textbf{Ref.}   \\ \hline
		\multirow{2}{*}{\rotatebox[origin=c]{90}{\textbf{Small}}} & Fermat & ($x_a$,$y_a$,$\psi_a$) & 0 & - &  Root Finding & F         \\
		& Fermat & ($x_c$,$y_c$,$\psi_c$) & 0 & - & Root Finding & T \\
		\hline
		\multirow{3}{*}{\rotatebox[origin=c]{90}{\textbf{Large}}} & Fermat & ($x_a$,$y_a$,$\psi_a$) & 0  & - & Eq. \eqref{eq:theta_kmax} & F         \\
		& arc      & ($x_t$,$y_t$,$\psi_t$) & $k_{max}$ & $\frac{\psi_{\Delta}-2\psi_{t}}{k_{max}}$ & - & -  \\
		& Fermat & ($x_{c}$,$y_{c}$,$\psi_{c}$) & 0 & - & Eq. \eqref{eq:theta_kmax} & T     
	\end{tabular}
\end{table}

The final parameter of interest in the Fermat Fillet method is the distance, $d$, to the connection point on the first straight segment. Similar to the approach in the Clothoid Fillet method, $d_{t}$ is computed using \eqref{eq:dt_connection} where $y_{m}$ is either the mid-point of the $k_{max}$ arc segment or $y_{t}$ if there is no $k_{max}$ segment. Then the distance $d$ is calculated as
\nolinebreak
\begin{eqnarray*}
d & = & x_{t}+d_{t}
\end{eqnarray*}
where $x_{t}$ is the $x$ value of the transition point between the two Fermat segments, or it is the $x$ value of the mid-point of the $k_{max}$ arc, if one is required.

Finally, the initial position and heading for each of the smoothing segments are adjusted based on the calculated attachment position and course angle. The adjustment is accomplished by adding the attachment position and course angle to the initial position and course angle of each of the segments.

%% file: figures/arc_fillet.tex
\tikzset{every picture/.style={line width=0.75pt}} %set default line width to 0.75pt        

\begin{tikzpicture}[x=0.75pt,y=0.75pt,yscale=-1,xscale=1]
%uncomment if require: \path (0,340); %set diagram left start at 0, and has height of 340

%Straight Lines [id:da7037058691998712] 
\draw [color={rgb, 255:red, 0; green, 0; blue, 0 }  ,draw opacity=1 ]   (112.86,240.71) -- (178.11,240.71) ;
\draw [shift={(180.11,240.71)}, rotate = 180] [color={rgb, 255:red, 0; green, 0; blue, 0 }  ,draw opacity=1 ][line width=0.75]    (10.93,-4.9) .. controls (6.95,-2.3) and (3.31,-0.67) .. (0,0) .. controls (3.31,0.67) and (6.95,2.3) .. (10.93,4.9)   ;
%Straight Lines [id:da46368583067430014] 
\draw [color={rgb, 255:red, 0; green, 0; blue, 0 }  ,draw opacity=1 ]   (112.86,240.71) -- (112.86,170.54) ;
\draw [shift={(112.86,168.54)}, rotate = 450] [color={rgb, 255:red, 0; green, 0; blue, 0 }  ,draw opacity=1 ][line width=0.75]    (10.93,-4.9) .. controls (6.95,-2.3) and (3.31,-0.67) .. (0,0) .. controls (3.31,0.67) and (6.95,2.3) .. (10.93,4.9)   ;
%Straight Lines [id:da2373759738816732] 
\draw [color={rgb, 255:red, 0; green, 0; blue, 0 }  ,draw opacity=1 ] [dash pattern={on 4.5pt off 4.5pt}]  (112.86,240.71) -- (180.11,240.71) -- (241.39,240.35) ;
%Straight Lines [id:da6747251631647537] 
\draw  [dash pattern={on 0.84pt off 2.51pt}]  (241.39,240.35) -- (280.1,240.4) ;
%Shape: Arc [id:dp7747406168697106] 
\draw  [draw opacity=0] (227.47,221.03) .. controls (231.29,217.69) and (236.13,215.7) .. (241.39,215.7) .. controls (253.16,215.7) and (262.81,225.68) .. (263.74,238.36) -- (241.39,240.35) -- cycle ; \draw   (227.47,221.03) .. controls (231.29,217.69) and (236.13,215.7) .. (241.39,215.7) .. controls (253.16,215.7) and (262.81,225.68) .. (263.74,238.36) ;
%Straight Lines [id:da5471101882302769] 
\draw [color={rgb, 255:red, 0; green, 0; blue, 0 }  ,draw opacity=1 ] [dash pattern={on 4.5pt off 4.5pt}]  (134.1,111.4) -- (241.39,240.35) ;
%Straight Lines [id:da6287138362700497] 
\draw    (116.53,274.96) -- (240.1,274.4) ;
%Straight Lines [id:da7912117123792477] 
\draw    (116.53,262.93) -- (116.53,286.99) ;
%Straight Lines [id:da9781397060011314] 
\draw    (240.1,262.37) -- (240.1,286.43) ;
%Straight Lines [id:da9081694523769044] 
\draw    (170.62,261.96) -- (170.62,274.96) ;
%Shape: Arc [id:dp11080833128327083] 
\draw  [draw opacity=0][line width=1.5]  (162.33,146.01) .. controls (170,155.92) and (174.55,168.3) .. (174.55,181.71) .. controls (174.55,213.98) and (148.22,240.19) .. (115.53,240.71) -- (114.55,181.71) -- cycle ; \draw  [color={rgb, 255:red, 208; green, 2; blue, 27 }  ,draw opacity=1 ][line width=1.5]  (162.33,146.01) .. controls (170,155.92) and (174.55,168.3) .. (174.55,181.71) .. controls (174.55,213.98) and (148.22,240.19) .. (115.53,240.71) ;
%Straight Lines [id:da17612530441903562] 
\draw [color={rgb, 255:red, 144; green, 19; blue, 254 }  ,draw opacity=1 ][line width=1.5]    (134.1,111.4) -- (162.33,146.01) ;
%Flowchart: Connector [id:dp9251384575011634] 
\draw  [fill={rgb, 255:red, 0; green, 0; blue, 0 }  ,fill opacity=1 ] (159.66,146.01) .. controls (159.66,144.42) and (160.85,143.13) .. (162.33,143.13) .. controls (163.8,143.13) and (164.99,144.42) .. (164.99,146.01) .. controls (164.99,147.61) and (163.8,148.9) .. (162.33,148.9) .. controls (160.85,148.9) and (159.66,147.61) .. (159.66,146.01) -- cycle ;
%Flowchart: Connector [id:dp24098251469586973] 
\draw  [fill={rgb, 255:red, 0; green, 0; blue, 0 }  ,fill opacity=1 ] (110.2,240.71) .. controls (110.2,239.11) and (111.39,237.82) .. (112.86,237.82) .. controls (114.34,237.82) and (115.53,239.11) .. (115.53,240.71) .. controls (115.53,242.3) and (114.34,243.59) .. (112.86,243.59) .. controls (111.39,243.59) and (110.2,242.3) .. (110.2,240.71) -- cycle ;
%Flowchart: Connector [id:dp24920273333021248] 
\draw  [fill={rgb, 255:red, 0; green, 0; blue, 0 }  ,fill opacity=1 ] (238.72,240.35) .. controls (238.72,238.76) and (239.92,237.47) .. (241.39,237.47) .. controls (242.86,237.47) and (244.06,238.76) .. (244.06,240.35) .. controls (244.06,241.95) and (242.86,243.24) .. (241.39,243.24) .. controls (239.92,243.24) and (238.72,241.95) .. (238.72,240.35) -- cycle ;
%Flowchart: Connector [id:dp3283718034707179] 
\draw  [fill={rgb, 255:red, 0; green, 0; blue, 0 }  ,fill opacity=1 ] (131.43,111.4) .. controls (131.43,109.81) and (132.63,108.52) .. (134.1,108.52) .. controls (135.57,108.52) and (136.77,109.81) .. (136.77,111.4) .. controls (136.77,112.99) and (135.57,114.28) .. (134.1,114.28) .. controls (132.63,114.28) and (131.43,112.99) .. (131.43,111.4) -- cycle ;
%Flowchart: Connector [id:dp9342799311574037] 
\draw  [fill={rgb, 255:red, 0; green, 0; blue, 0 }  ,fill opacity=1 ] (166.2,206.71) .. controls (166.2,205.11) and (167.39,203.82) .. (168.86,203.82) .. controls (170.34,203.82) and (171.53,205.11) .. (171.53,206.71) .. controls (171.53,208.3) and (170.34,209.59) .. (168.86,209.59) .. controls (167.39,209.59) and (166.2,208.3) .. (166.2,206.71) -- cycle ;

% Text Node
\draw (131.55,246.66) node  [font=\small,color={rgb, 255:red, 0; green, 0; blue, 0 }  ,opacity=1 ]  {$x$};
% Text Node
\draw (106.96,223.77) node  [font=\small,color={rgb, 255:red, 0; green, 0; blue, 0 }  ,opacity=1 ]  {$y$};
% Text Node
\draw (262.5,206.5) node    {$\psi _{\Delta }$};
% Text Node
\draw (173.86,285.59) node    {$d$};
% Text Node
\draw (145.86,262.59) node    {$x_{m}$};
% Text Node
\draw (202.86,263.59) node    {$d_{t}$};
% Text Node
\draw (142,194.4) node  [font=\small]  {$( x_{m} ,y_{m})$};
% Text Node
\draw (193,131.4) node  [font=\small]  {$( x_{t} ,y_{t})$};

\end{tikzpicture}

%% file: figures/clothoid_segments.tex
\tikzset{every picture/.style={line width=0.75pt}} %set default line width to 0.75pt        

\begin{tikzpicture}[x=0.75pt,y=0.75pt,yscale=-1,xscale=1]
%uncomment if require: \path (0,340); %set diagram left start at 0, and has height of 340

%Curve Lines [id:da38552371336404057] 
\draw [color={rgb, 255:red, 144; green, 19; blue, 254 }  ,draw opacity=1 ][line width=1.5]    (95.53,220.14) .. controls (124,220.62) and (171.08,201.38) .. (189.69,188.49) ;
%Curve Lines [id:da4664188123111326] 
\draw [color={rgb, 255:red, 144; green, 19; blue, 254 }  ,draw opacity=1 ][line width=1.5]    (154.94,57.77) .. controls (176.5,77.57) and (197.5,124.57) .. (203.38,143.56) ;
%Straight Lines [id:da6341609088022722] 
\draw [color={rgb, 255:red, 0; green, 0; blue, 0 }  ,draw opacity=1 ]   (100.87,220.14) -- (166.12,220.14) ;
\draw [shift={(168.12,220.14)}, rotate = 180] [color={rgb, 255:red, 0; green, 0; blue, 0 }  ,draw opacity=1 ][line width=0.75]    (10.93,-4.9) .. controls (6.95,-2.3) and (3.31,-0.67) .. (0,0) .. controls (3.31,0.67) and (6.95,2.3) .. (10.93,4.9)   ;
%Straight Lines [id:da9087576499901888] 
\draw [color={rgb, 255:red, 0; green, 0; blue, 0 }  ,draw opacity=1 ]   (95.53,220.14) -- (95.53,149.98) ;
\draw [shift={(95.53,147.98)}, rotate = 450] [color={rgb, 255:red, 0; green, 0; blue, 0 }  ,draw opacity=1 ][line width=0.75]    (10.93,-4.9) .. controls (6.95,-2.3) and (3.31,-0.67) .. (0,0) .. controls (3.31,0.67) and (6.95,2.3) .. (10.93,4.9)   ;
%Straight Lines [id:da0748360420096934] 
\draw [color={rgb, 255:red, 0; green, 0; blue, 0 }  ,draw opacity=1 ] [dash pattern={on 4.5pt off 4.5pt}]  (95.53,220.14) -- (320.17,220.33) ;
%Flowchart: Connector [id:dp5863733768614277] 
\draw  [fill={rgb, 255:red, 0; green, 0; blue, 0 }  ,fill opacity=1 ] (152.27,57.77) .. controls (152.27,56.18) and (153.46,54.88) .. (154.94,54.88) .. controls (156.41,54.88) and (157.6,56.18) .. (157.6,57.77) .. controls (157.6,59.36) and (156.41,60.65) .. (154.94,60.65) .. controls (153.46,60.65) and (152.27,59.36) .. (152.27,57.77) -- cycle ;
%Straight Lines [id:da33984636613116526] 
\draw  [dash pattern={on 0.84pt off 2.51pt}]  (320.17,220.33) -- (380,220) ;
%Shape: Arc [id:dp3896803675833931] 
\draw  [draw opacity=0] (304.66,202.02) .. controls (308.48,198.68) and (313.32,196.69) .. (318.58,196.69) .. controls (330.35,196.69) and (340,206.67) .. (340.92,219.35) -- (318.58,221.34) -- cycle ; \draw   (304.66,202.02) .. controls (308.48,198.68) and (313.32,196.69) .. (318.58,196.69) .. controls (330.35,196.69) and (340,206.67) .. (340.92,219.35) ;
%Straight Lines [id:da2547857423216895] 
\draw [color={rgb, 255:red, 144; green, 19; blue, 254 }  ,draw opacity=1 ] [dash pattern={on 0.84pt off 2.51pt}]  (203.79,170.71) -- (320.17,220.33) ;
%Shape: Arc [id:dp019088423612496852] 
\draw  [draw opacity=0][line width=1.5]  (203.61,144.15) .. controls (207.18,153.78) and (206.87,165.01) .. (201.87,174.84) .. controls (198.99,180.51) and (194.9,185.01) .. (190.14,188.2) -- (172.45,157.6) -- cycle ; \draw  [color={rgb, 255:red, 208; green, 2; blue, 27 }  ,draw opacity=1 ][line width=1.5]  (203.61,144.15) .. controls (207.18,153.78) and (206.87,165.01) .. (201.87,174.84) .. controls (198.99,180.51) and (194.9,185.01) .. (190.14,188.2) ;
%Straight Lines [id:da570944185387708] 
\draw [color={rgb, 255:red, 0; green, 0; blue, 0 }  ,draw opacity=1 ] [dash pattern={on 4.5pt off 4.5pt}]  (154.94,57.77) -- (319.7,217.74) ;
%Straight Lines [id:da2541034716316979] 
\draw    (96.53,257.28) -- (320.7,257.28) ;
%Straight Lines [id:da2147231251549797] 
\draw    (96.53,245.25) -- (96.53,269.31) ;
%Straight Lines [id:da558200998325918] 
\draw    (320.7,245.25) -- (320.7,269.31) ;
%Shape: Arc [id:dp45779411581284046] 
\draw  [draw opacity=0] (269.31,219.78) .. controls (269.28,219.1) and (269.26,218.42) .. (269.26,217.74) .. controls (269.26,212.27) and (270.22,207.01) .. (271.98,202.11) -- (319.7,217.74) -- cycle ; \draw   (269.31,219.78) .. controls (269.28,219.1) and (269.26,218.42) .. (269.26,217.74) .. controls (269.26,212.27) and (270.22,207.01) .. (271.98,202.11) ;
%Flowchart: Connector [id:dp5117013270939892] 
\draw  [fill={rgb, 255:red, 0; green, 0; blue, 0 }  ,fill opacity=1 ] (200.71,143.56) .. controls (200.71,141.96) and (201.91,140.67) .. (203.38,140.67) .. controls (204.86,140.67) and (206.05,141.96) .. (206.05,143.56) .. controls (206.05,145.15) and (204.86,146.44) .. (203.38,146.44) .. controls (201.91,146.44) and (200.71,145.15) .. (200.71,143.56) -- cycle ;
%Flowchart: Connector [id:dp3282145031444401] 
\draw  [fill={rgb, 255:red, 0; green, 0; blue, 0 }  ,fill opacity=1 ] (187.48,188.2) .. controls (187.48,186.6) and (188.67,185.31) .. (190.14,185.31) .. controls (191.62,185.31) and (192.81,186.6) .. (192.81,188.2) .. controls (192.81,189.79) and (191.62,191.08) .. (190.14,191.08) .. controls (188.67,191.08) and (187.48,189.79) .. (187.48,188.2) -- cycle ;
%Flowchart: Connector [id:dp7794678283907222] 
\draw  [fill={rgb, 255:red, 0; green, 0; blue, 0 }  ,fill opacity=1 ] (92.86,220.14) .. controls (92.86,218.55) and (94.06,217.26) .. (95.53,217.26) .. controls (97,217.26) and (98.2,218.55) .. (98.2,220.14) .. controls (98.2,221.73) and (97,223.03) .. (95.53,223.03) .. controls (94.06,223.03) and (92.86,221.73) .. (92.86,220.14) -- cycle ;
%Flowchart: Connector [id:dp436091172779387] 
\draw  [fill={rgb, 255:red, 0; green, 0; blue, 0 }  ,fill opacity=1 ] (201.12,170.71) .. controls (201.12,169.12) and (202.31,167.83) .. (203.79,167.83) .. controls (205.26,167.83) and (206.45,169.12) .. (206.45,170.71) .. controls (206.45,172.31) and (205.26,173.6) .. (203.79,173.6) .. controls (202.31,173.6) and (201.12,172.31) .. (201.12,170.71) -- cycle ;
%Flowchart: Connector [id:dp2351300459521497] 
\draw  [fill={rgb, 255:red, 0; green, 0; blue, 0 }  ,fill opacity=1 ] (320.17,220.33) .. controls (320.17,218.74) and (321.36,217.45) .. (322.83,217.45) .. controls (324.31,217.45) and (325.5,218.74) .. (325.5,220.33) .. controls (325.5,221.93) and (324.31,223.22) .. (322.83,223.22) .. controls (321.36,223.22) and (320.17,221.93) .. (320.17,220.33) -- cycle ;
%Straight Lines [id:da8316872134808111] 
\draw    (203.79,170.71) -- (214.35,133.72) ;
\draw [shift={(214.9,131.8)}, rotate = 465.94] [color={rgb, 255:red, 0; green, 0; blue, 0 }  ][line width=0.75]    (10.93,-3.29) .. controls (6.95,-1.4) and (3.31,-0.3) .. (0,0) .. controls (3.31,0.3) and (6.95,1.4) .. (10.93,3.29)   ;
%Curve Lines [id:da4802253051716161] 
\draw    (208.1,161.2) .. controls (214,161.25) and (216.16,164.84) .. (217.1,169.2) ;
%Straight Lines [id:da10289921742420916] 
\draw    (189.69,188.49) -- (211.45,173.72) ;
\draw [shift={(213.1,172.6)}, rotate = 505.82] [color={rgb, 255:red, 0; green, 0; blue, 0 }  ][line width=0.75]    (10.93,-3.29) .. controls (6.95,-1.4) and (3.31,-0.3) .. (0,0) .. controls (3.31,0.3) and (6.95,1.4) .. (10.93,3.29)   ;
%Curve Lines [id:da6588963265358052] 
\draw    (196.7,184) .. controls (200.3,183) and (200.3,185) .. (200.7,188) ;
%Straight Lines [id:da4681063942575945] 
\draw  [dash pattern={on 0.84pt off 2.51pt}]  (203.79,170.71) -- (253.1,170.6) ;
%Straight Lines [id:da7934562273350625] 
\draw  [dash pattern={on 0.84pt off 2.51pt}]  (192.81,188.2) -- (228.7,188) ;
%Straight Lines [id:da9251702912030713] 
\draw    (203.5,244.5) -- (203.5,257.5) ;

% Text Node
\draw (114.55,226.98) node  [font=\small,color={rgb, 255:red, 0; green, 0; blue, 0 }  ,opacity=1 ]  {$x$};
% Text Node
\draw (87.96,202.09) node  [font=\small,color={rgb, 255:red, 0; green, 0; blue, 0 }  ,opacity=1 ]  {$y$};
% Text Node
\draw (338.81,184.63) node    {$\psi _{\Delta }$};
% Text Node
\draw (248.69,124.5) node    {$d$};
% Text Node
\draw (209.86,271.91) node    {$d$};
% Text Node
\draw (255.07,207.49) node    {$\phi $};
% Text Node
\draw (162,179) node    {$( x_{t} ,y_{t})$};
% Text Node
\draw (231.81,151.63) node    {$\frac{\psi _{\Delta }}{2}$};
% Text Node
\draw (207.81,194.63) node    {$\psi _{t}$};
% Text Node
\draw (151.86,246.91) node    {$x_{t}$};
% Text Node
\draw (255.86,244.91) node    {$d_{t}$};

\end{tikzpicture}

%% file: figures/arc_calc.tex
\tikzset{every picture/.style={line width=0.75pt}} %set default line width to 0.75pt        

\begin{tikzpicture}[x=0.75pt,y=0.75pt,yscale=-1,xscale=1]
%uncomment if require: \path (0,340); %set diagram left start at 0, and has height of 340

%Curve Lines [id:da6240134778092274] 
\draw    (413.1,220.8) .. controls (418.9,220.8) and (421.1,224.4) .. (417.1,228.8) ;
%Straight Lines [id:da7817596840767196] 
\draw    (403.71,228.82) -- (448.78,190.3) ;
\draw [shift={(450.3,189)}, rotate = 499.47] [color={rgb, 255:red, 0; green, 0; blue, 0 }  ][line width=0.75]    (10.93,-3.29) .. controls (6.95,-1.4) and (3.31,-0.3) .. (0,0) .. controls (3.31,0.3) and (6.95,1.4) .. (10.93,3.29)   ;
%Shape: Ellipse [id:dp9296749949487197] 
\draw  [dash pattern={on 0.84pt off 2.51pt}] (273.48,168.68) .. controls (273.48,126.33) and (309.34,92) .. (353.59,92) .. controls (397.83,92) and (433.7,126.33) .. (433.7,168.68) .. controls (433.7,211.03) and (397.83,245.36) .. (353.59,245.36) .. controls (309.34,245.36) and (273.48,211.03) .. (273.48,168.68) -- cycle ;
%Shape: Arc [id:dp8199525754926991] 
\draw  [draw opacity=0][line width=1.5]  (403.1,108.51) .. controls (421.74,122.59) and (433.7,144.41) .. (433.7,168.9) .. controls (433.7,193.12) and (422,214.73) .. (403.71,228.82) -- (353.36,168.9) -- cycle ; \draw  [color={rgb, 255:red, 208; green, 2; blue, 27 }  ,draw opacity=1 ][line width=1.5]  (403.1,108.51) .. controls (421.74,122.59) and (433.7,144.41) .. (433.7,168.9) .. controls (433.7,193.12) and (422,214.73) .. (403.71,228.82) ;
%Straight Lines [id:da6304822122882474] 
\draw    (353.59,168.68) -- (403.71,228.82) ;
%Straight Lines [id:da7427607025342078] 
\draw    (353.36,168.9) -- (403.1,108.51) ;
%Curve Lines [id:da2854459287670028] 
\draw    (362.81,158.39) .. controls (375.05,158.39) and (378.11,175.97) .. (364.34,178.9) ;
%Straight Lines [id:da4432329501533796] 
\draw    (433.7,168.68) -- (433.31,117) ;
\draw [shift={(433.3,115)}, rotate = 449.57] [color={rgb, 255:red, 0; green, 0; blue, 0 }  ][line width=0.75]    (10.93,-3.29) .. controls (6.95,-1.4) and (3.31,-0.3) .. (0,0) .. controls (3.31,0.3) and (6.95,1.4) .. (10.93,3.29)   ;
%Straight Lines [id:da9011972581639229] 
\draw    (403.1,108.51) -- (378.96,85.97) ;
\draw [shift={(377.5,84.6)}, rotate = 403.03999999999996] [color={rgb, 255:red, 0; green, 0; blue, 0 }  ][line width=0.75]    (10.93,-3.29) .. controls (6.95,-1.4) and (3.31,-0.3) .. (0,0) .. controls (3.31,0.3) and (6.95,1.4) .. (10.93,3.29)   ;
%Straight Lines [id:da9626974463294016] 
\draw  [dash pattern={on 0.84pt off 2.51pt}]  (403.71,228.82) -- (458.3,228.6) ;
%Straight Lines [id:da14703848189971858] 
\draw  [dash pattern={on 0.84pt off 2.51pt}]  (433.7,168.68) -- (488.29,168.46) ;
%Straight Lines [id:da8112519636473001] 
\draw  [dash pattern={on 0.84pt off 2.51pt}]  (403.1,108.51) -- (457.69,108.29) ;
%Curve Lines [id:da3356835181760529] 
\draw    (435.1,160.8) .. controls (440.9,160.8) and (443.1,164.4) .. (442.1,168.8) ;
%Curve Lines [id:da840095807710384] 
\draw    (398.1,102.8) .. controls (403.9,99.8) and (412.1,101.8) .. (413.1,108.8) ;
%Shape: Circle [id:dp5035702489629736] 
\draw  [fill={rgb, 255:red, 0; green, 0; blue, 0 }  ,fill opacity=1 ] (400.71,228.82) .. controls (400.71,227.17) and (402.06,225.82) .. (403.71,225.82) .. controls (405.37,225.82) and (406.71,227.17) .. (406.71,228.82) .. controls (406.71,230.48) and (405.37,231.82) .. (403.71,231.82) .. controls (402.06,231.82) and (400.71,230.48) .. (400.71,228.82) -- cycle ;
%Shape: Circle [id:dp9686986282800927] 
\draw  [fill={rgb, 255:red, 0; green, 0; blue, 0 }  ,fill opacity=1 ] (350.59,168.68) .. controls (350.59,167.02) and (351.93,165.68) .. (353.59,165.68) .. controls (355.25,165.68) and (356.59,167.02) .. (356.59,168.68) .. controls (356.59,170.34) and (355.25,171.68) .. (353.59,171.68) .. controls (351.93,171.68) and (350.59,170.34) .. (350.59,168.68) -- cycle ;
%Shape: Circle [id:dp2603392538213647] 
\draw  [fill={rgb, 255:red, 0; green, 0; blue, 0 }  ,fill opacity=1 ] (430.7,168.68) .. controls (430.7,167.02) and (432.04,165.68) .. (433.7,165.68) .. controls (435.36,165.68) and (436.7,167.02) .. (436.7,168.68) .. controls (436.7,170.34) and (435.36,171.68) .. (433.7,171.68) .. controls (432.04,171.68) and (430.7,170.34) .. (430.7,168.68) -- cycle ;
%Shape: Circle [id:dp02394363944934108] 
\draw  [fill={rgb, 255:red, 0; green, 0; blue, 0 }  ,fill opacity=1 ] (400.1,108.51) .. controls (400.1,106.85) and (401.45,105.51) .. (403.1,105.51) .. controls (404.76,105.51) and (406.1,106.85) .. (406.1,108.51) .. controls (406.1,110.17) and (404.76,111.51) .. (403.1,111.51) .. controls (401.45,111.51) and (400.1,110.17) .. (400.1,108.51) -- cycle ;
%Straight Lines [id:da2111004507262566] 
\draw  [dash pattern={on 0.84pt off 2.51pt}]  (353.59,168.68) -- (354.7,228) ;
%Straight Lines [id:da31851194279730977] 
\draw  [dash pattern={on 0.84pt off 2.51pt}]  (354.7,228) -- (403.71,228.82) ;
%Shape: Right Angle [id:dp8290745958773662] 
\draw   (354.7,220.59) -- (362.2,220.59) -- (362.2,228) ;

% Text Node
\draw (383.16,164.39) node    {$\phi $};
% Text Node
\draw (363,121.8) node    {$\frac{1}{k_{max}}$};
% Text Node
\draw (436,216.8) node    {$\psi _{t_1}$};
% Text Node
\draw (453,148.8) node    {$\frac{\psi _{\Delta }}{2}$};
% Text Node
\draw (419,244.4) node    {$( x_{t_1} ,y_{t_1})$};
% Text Node
\draw (370,234.2) node  [font=\footnotesize]  {$dx$};
% Text Node
\draw (345,196.2) node  [font=\footnotesize]  {$dy$};
% Text Node
\draw (327,154.4) node    {$( x_{c} ,y_{c})$};
% Text Node
\draw (432,83.4) node    {$( x_{t_{2}} ,y_{t_{2}})$};

\end{tikzpicture}

%% file: figures/fermat_fillet.tex
\tikzset{every picture/.style={line width=0.75pt}} %set default line width to 0.75pt        

\begin{tikzpicture}[x=0.75pt,y=0.75pt,yscale=-1,xscale=1]
%uncomment if require: \path (0,340); %set diagram left start at 0, and has height of 340

%Curve Lines [id:da38552371336404057] 
\draw [color={rgb, 255:red, 144; green, 19; blue, 254 }  ,draw opacity=1 ][line width=1.5]    (95.53,220.14) .. controls (124,220.62) and (171.08,201.38) .. (189.69,188.49) ;
%Curve Lines [id:da4664188123111326] 
\draw [color={rgb, 255:red, 144; green, 19; blue, 254 }  ,draw opacity=1 ][line width=1.5]    (154.94,57.77) .. controls (176.5,77.57) and (197.5,124.57) .. (203.38,143.56) ;
%Straight Lines [id:da6341609088022722] 
\draw [color={rgb, 255:red, 0; green, 0; blue, 0 }  ,draw opacity=1 ]   (100.87,220.14) -- (166.12,220.14) ;
\draw [shift={(168.12,220.14)}, rotate = 180] [color={rgb, 255:red, 0; green, 0; blue, 0 }  ,draw opacity=1 ][line width=0.75]    (10.93,-4.9) .. controls (6.95,-2.3) and (3.31,-0.67) .. (0,0) .. controls (3.31,0.67) and (6.95,2.3) .. (10.93,4.9)   ;
%Straight Lines [id:da9087576499901888] 
\draw [color={rgb, 255:red, 0; green, 0; blue, 0 }  ,draw opacity=1 ]   (95.53,220.14) -- (95.53,149.98) ;
\draw [shift={(95.53,147.98)}, rotate = 450] [color={rgb, 255:red, 0; green, 0; blue, 0 }  ,draw opacity=1 ][line width=0.75]    (10.93,-4.9) .. controls (6.95,-2.3) and (3.31,-0.67) .. (0,0) .. controls (3.31,0.67) and (6.95,2.3) .. (10.93,4.9)   ;
%Straight Lines [id:da0748360420096934] 
\draw [color={rgb, 255:red, 0; green, 0; blue, 0 }  ,draw opacity=1 ] [dash pattern={on 4.5pt off 4.5pt}]  (95.53,220.14) -- (320.17,220.33) ;
%Flowchart: Connector [id:dp5863733768614277] 
\draw  [fill={rgb, 255:red, 0; green, 0; blue, 0 }  ,fill opacity=1 ] (152.27,57.77) .. controls (152.27,56.18) and (153.46,54.88) .. (154.94,54.88) .. controls (156.41,54.88) and (157.6,56.18) .. (157.6,57.77) .. controls (157.6,59.36) and (156.41,60.65) .. (154.94,60.65) .. controls (153.46,60.65) and (152.27,59.36) .. (152.27,57.77) -- cycle ;
%Straight Lines [id:da33984636613116526] 
\draw  [dash pattern={on 0.84pt off 2.51pt}]  (320.17,220.33) -- (380,220) ;
%Shape: Arc [id:dp3896803675833931] 
\draw  [draw opacity=0] (304.66,202.02) .. controls (308.48,198.68) and (313.32,196.69) .. (318.58,196.69) .. controls (330.35,196.69) and (340,206.67) .. (340.92,219.35) -- (318.58,221.34) -- cycle ; \draw   (304.66,202.02) .. controls (308.48,198.68) and (313.32,196.69) .. (318.58,196.69) .. controls (330.35,196.69) and (340,206.67) .. (340.92,219.35) ;
%Shape: Arc [id:dp019088423612496852] 
\draw  [draw opacity=0][line width=1.5]  (203.61,144.15) .. controls (207.18,153.78) and (206.87,165.01) .. (201.87,174.84) .. controls (198.99,180.51) and (194.9,185.01) .. (190.14,188.2) -- (172.45,157.6) -- cycle ; \draw  [color={rgb, 255:red, 208; green, 2; blue, 27 }  ,draw opacity=1 ][line width=1.5]  (203.61,144.15) .. controls (207.18,153.78) and (206.87,165.01) .. (201.87,174.84) .. controls (198.99,180.51) and (194.9,185.01) .. (190.14,188.2) ;
%Straight Lines [id:da570944185387708] 
\draw [color={rgb, 255:red, 0; green, 0; blue, 0 }  ,draw opacity=1 ] [dash pattern={on 4.5pt off 4.5pt}]  (154.94,57.77) -- (319.7,217.74) ;
%Straight Lines [id:da2541034716316979] 
\draw    (96.53,257.28) -- (320.7,257.28) ;
%Straight Lines [id:da2147231251549797] 
\draw    (96.53,245.25) -- (96.53,269.31) ;
%Straight Lines [id:da558200998325918] 
\draw    (320.7,245.25) -- (320.7,269.31) ;
%Flowchart: Connector [id:dp5117013270939892] 
\draw  [fill={rgb, 255:red, 0; green, 0; blue, 0 }  ,fill opacity=1 ] (200.71,143.56) .. controls (200.71,141.96) and (201.91,140.67) .. (203.38,140.67) .. controls (204.86,140.67) and (206.05,141.96) .. (206.05,143.56) .. controls (206.05,145.15) and (204.86,146.44) .. (203.38,146.44) .. controls (201.91,146.44) and (200.71,145.15) .. (200.71,143.56) -- cycle ;
%Flowchart: Connector [id:dp3282145031444401] 
\draw  [fill={rgb, 255:red, 0; green, 0; blue, 0 }  ,fill opacity=1 ] (187.48,188.2) .. controls (187.48,186.6) and (188.67,185.31) .. (190.14,185.31) .. controls (191.62,185.31) and (192.81,186.6) .. (192.81,188.2) .. controls (192.81,189.79) and (191.62,191.08) .. (190.14,191.08) .. controls (188.67,191.08) and (187.48,189.79) .. (187.48,188.2) -- cycle ;
%Flowchart: Connector [id:dp7794678283907222] 
\draw  [fill={rgb, 255:red, 0; green, 0; blue, 0 }  ,fill opacity=1 ] (92.86,220.14) .. controls (92.86,218.55) and (94.06,217.26) .. (95.53,217.26) .. controls (97,217.26) and (98.2,218.55) .. (98.2,220.14) .. controls (98.2,221.73) and (97,223.03) .. (95.53,223.03) .. controls (94.06,223.03) and (92.86,221.73) .. (92.86,220.14) -- cycle ;
%Flowchart: Connector [id:dp436091172779387] 
\draw  [fill={rgb, 255:red, 0; green, 0; blue, 0 }  ,fill opacity=1 ] (201.12,170.71) .. controls (201.12,169.12) and (202.31,167.83) .. (203.79,167.83) .. controls (205.26,167.83) and (206.45,169.12) .. (206.45,170.71) .. controls (206.45,172.31) and (205.26,173.6) .. (203.79,173.6) .. controls (202.31,173.6) and (201.12,172.31) .. (201.12,170.71) -- cycle ;
%Flowchart: Connector [id:dp2351300459521497] 
\draw  [fill={rgb, 255:red, 0; green, 0; blue, 0 }  ,fill opacity=1 ] (320.17,220.33) .. controls (320.17,218.74) and (321.36,217.45) .. (322.83,217.45) .. controls (324.31,217.45) and (325.5,218.74) .. (325.5,220.33) .. controls (325.5,221.93) and (324.31,223.22) .. (322.83,223.22) .. controls (321.36,223.22) and (320.17,221.93) .. (320.17,220.33) -- cycle ;
%Straight Lines [id:da9251702912030713] 
\draw    (203.5,244.5) -- (203.5,257.5) ;
%Straight Lines [id:da02190322220202412] 
\draw  [dash pattern={on 0.84pt off 2.51pt}]  (203.38,143.56) -- (223.5,125) ;
%Straight Lines [id:da9614363750492625] 
\draw  [dash pattern={on 0.84pt off 2.51pt}]  (154.94,57.77) -- (154.5,144) ;
%Straight Lines [id:da5732418234982126] 
\draw  [dash pattern={on 0.84pt off 2.51pt}]  (203.38,143.56) -- (154.5,144) ;
%Straight Lines [id:da9274134460904302] 
\draw  [dash pattern={on 0.84pt off 2.51pt}]  (154.94,57.77) -- (214.77,57.44) ;
%Shape: Arc [id:dp3812648694148759] 
\draw  [draw opacity=0] (177.34,56.91) .. controls (177.35,57.19) and (177.35,57.48) .. (177.35,57.77) .. controls (177.35,62.98) and (175.88,67.82) .. (173.37,71.8) -- (154.94,57.77) -- cycle ; \draw   (177.34,56.91) .. controls (177.35,57.19) and (177.35,57.48) .. (177.35,57.77) .. controls (177.35,62.98) and (175.88,67.82) .. (173.37,71.8) ;
%Shape: Arc [id:dp14640815160041676] 
\draw  [draw opacity=0] (165.51,79.52) .. controls (162.36,81.37) and (158.76,82.43) .. (154.94,82.43) .. controls (154.77,82.43) and (154.61,82.42) .. (154.44,82.42) -- (154.94,57.77) -- cycle ; \draw   (165.51,79.52) .. controls (162.36,81.37) and (158.76,82.43) .. (154.94,82.43) .. controls (154.77,82.43) and (154.61,82.42) .. (154.44,82.42) ;
%Straight Lines [id:da9123698091846253] 
\draw  [dash pattern={on 0.84pt off 2.51pt}]  (154.94,57.77) -- (203.38,143.56) ;

% Text Node
\draw (114.55,226.98) node  [font=\small,color={rgb, 255:red, 0; green, 0; blue, 0 }  ,opacity=1 ]  {$x$};
% Text Node
\draw (87.96,202.09) node  [font=\small,color={rgb, 255:red, 0; green, 0; blue, 0 }  ,opacity=1 ]  {$y$};
% Text Node
\draw (338.81,184.63) node    {$\psi _{\Delta }$};
% Text Node
\draw (209.86,271.91) node    {$d$};
% Text Node
\draw (230,155) node  [rotate=-0.03]  {$( x_{t} ,y_{t})$};
% Text Node
\draw (151.86,246.91) node    {$x_{t}$};
% Text Node
\draw (255.86,244.91) node    {$d_{t}$};
% Text Node
\draw (204,117.4) node [anchor=north west][inner sep=0.75pt]    {$h$};
% Text Node
\draw (203,83.4) node [anchor=north west][inner sep=0.75pt]    {$l$};
% Text Node
\draw (121,68) node    {$( x_{c} ,y_{c})$};
% Text Node
\draw (189.81,67.63) node    {$\psi _{c}$};
% Text Node
\draw (156.44,85.82) node [anchor=north west][inner sep=0.75pt]    {$\phi $};
% Text Node
\draw (237,182) node  [rotate=-0.03]  {$( x_{m} ,y_{m})$};

\end{tikzpicture}

%% file: sections/ImuSignalGeneration.tex
The purpose of this section is to present a method to efficiently determine the accelerations and angular rates experienced by an aircraft following a path smoothed using the methods described in Section \ref{sec:Path-Smoothing}. The accelerations and angular rates can then be used as nominal measurements from an IMU and be used to propagate the covariance of the state estimate of a navigation system. This section will present an approach for determining the acceleration and angular rates for an aircraft using curvilinear motion theory (see Section \ref{sec:background_curv}), vehicle dynamics, and knowledge of the aircraft maneuver.

Other methods for generating accelerations and angular rates exist, namely, inverting the navigation equations, or 6-DOF aircraft simulations. These methods have potential to be highly accurate, but they are computationally expensive which may be prohibitive depending on the application. The method presented in this section is intended to efficiently provide accelerations and angular rates at high enough fidelity to provide accurate covariance propagation results. 

%\subsection{Accelerations}

The accelerations experienced by the aircraft in the NED frame can be expressed using curvilinear motion theory as 
\begin{eqnarray}
\boldsymbol{a}^{n} & = & \ddot{s}\boldsymbol{e}_{t}^{n}+\dot{s}\dot{\psi}\boldsymbol{e}_{n}^{n}\label{eq:curvilinear_accel}
\end{eqnarray}
where tangent vector in the 2D plane is computed as a function of the course angle as
\begin{eqnarray}
\boldsymbol{e}_{t}^{n} & = & \left[\begin{array}{ccc}
\cos\text{\ensuremath{\left(\psi\right)}} & \sin\left(\psi\right) & 0\end{array}\right]^{\intercal} \label{eq:tangent_vector}
\end{eqnarray}
and the normal vector is defined to complete the right-handed coordinate frame between the tangent vector and positive-z unit vector as  
\begin{eqnarray}
\boldsymbol{e}_{n}^{n} & = & \left[\begin{array}{ccc}
0 & 0 & 1\end{array}\right]^{\intercal}\times\boldsymbol{e}_{t}^{n}. \label{eq:normal_vector}
\end{eqnarray}
Combining \eqref{eq:curvilinear_accel} - \eqref{eq:normal_vector} results in
\begin{eqnarray}
\boldsymbol{a}^{n} & = & \begin{bmatrix} 
\ddot{s}\cos(\psi) - \dot{s}\dot{\psi}\sin(\psi) \\ 
\ddot{s}\sin(\psi) + \dot{s}\dot{\psi}\cos(\psi) \\
0 \\
\end{bmatrix} \label{eq:curvilinear_accel_final}
\end{eqnarray}
as the expression for the acceleration experienced by the aircraft in the NED frame and is valid for all of the segment types presented in this work.
%This expression is not dependent on knowledge of the vehicle maneuver which is not the case for the angular rates as will be shown in the following paragraphs.

%\subsection{Angular Rates}

The angular rates experienced by an aircraft are dependent on the path shape and the dynamics of the vehicle maneuver. An aircraft, for example, can perform a variety of maneuvers to negotiate a turn (i.e. skid-to-steer, coordinated turn) and each will result in different angular rates. In this work, the angular rates for an aircraft executing a coordinated turn will be examined. The coordinated turn is a common aircraft turning condition where there is no lateral acceleration in the body frame of the aircraft \cite{bishop_coordinated_turn}.

The coordinated turn assumption provides a useful relationship between the course angle $(\psi)$ and roll angle $(\phi)$ \cite{beard_randy_small_2012} of the aircraft as 
\begin{equation}
\dot{\psi}=\frac{g}{V_{a}}\tan\phi.
\label{eq:coordinated_turn_roll_rate}
\end{equation}
The rate of change of the roll angle can be computed by first differentiating \eqref{eq:coordinated_turn_roll_rate} as
\begin{equation}
\ddot{\psi}=\frac{g}{V_{a}}\frac{1}{\cos^{2}\phi}\dot{\phi}
\end{equation}
and solving for the roll angle rate, which yields
\begin{equation}
\dot{\phi}=\ddot{\psi}\frac{V_{a}}{g}\cos^{2}\phi.\label{eq:phidot-inter}
\end{equation}

The roll rate $(\dot{\phi})$ and course angle rate $(\dot{\psi})$ need to be converted to body-frame angular rates, $(p,q,r)$. This conversion requires a frame change from the Euler angle frames to the body-frame which will be completed using a ZYX Euler angle sequence. The Euler angle rates are related to the body-frame angular rates \cite{beard_randy_small_2012} as 
\begin{equation}
\left[\begin{array}{c}
p\\
q\\
r
\end{array}\right]=\underset{A_{B}}{\underbrace{\left[\begin{array}{ccc}
1 & 0 & -\sin\theta\\
0 & \cos\phi & \sin\phi\cos\theta\\
0 & -\sin\phi & \cos\phi\cos\theta
\end{array}\right]}}\left[\begin{array}{c}
\dot{\phi}\\
\dot{\theta}\\
\dot{\psi}
\end{array}\right].\label{eq:pqr}
\end{equation}
Assuming no change in pitch angle, substitution of \eqref{eq:phidot-inter}
and \eqref{eq:psidot} into \eqref{eq:pqr} yields
\begin{equation}
\left[\begin{array}{c}
p\\
q\\
r
\end{array}\right]=A_{B}\left[\begin{array}{c}
\ddot{\psi}\frac{V_{a}}{g}\cos^{2}\phi\\
0\\
\dot{\psi}
\end{array}\right]\label{eq:pqr-1}
\end{equation}
or in component form
\begin{eqnarray}
p & = & \ddot{\psi}\frac{V_{a}}{g}\cos^{2}\phi-\dot{\psi}\sin\theta \\
q & = & \dot{\psi}\sin\phi\cos\theta \\
r & = & \dot{\psi}\cos\phi\cos\theta.
\end{eqnarray}
This indicates that to compute the angular rates of an aircraft executing
a coordinated turn, $\dot{\psi}$ and $\ddot{\psi}$ must be determined
for each of the segment type in the path.

The time-derivative of the course angle is computed from  \eqref{eq:clothoid_heading} utilizing the chain rule
\begin{eqnarray}
\dot{\psi}(t)& = & \frac{d}{dt}\psi\left(s\left(t\right)\right) \\
 & = & \frac{d\psi}{d s}\frac{d s}{d t}.\label{eq:psidot}
\end{eqnarray}
The second derivative of heading with respect to time is computed utilizing a combination
of multiplication rule and chain rule as
\begin{eqnarray}
\ddot{\psi}(t) & = & \frac{d}{d t}\psi^{\prime}(s(t))\dot{s}(t)\nonumber \\
& = & \frac{d\psi^{\prime}}{d s}\frac{d s}{d t}\dot{s}(t)+\psi^{\prime}\ddot{s}(t)\nonumber \\
& = & \psi^{\prime\prime}(s(t))\dot{s}(t)^{2}+\psi^{\prime}(t)\ddot{s}(t).\label{eq:psiddot}
\end{eqnarray}

To generate the angular rates for the aircraft, it remains to define $\dot{\psi}$ and $\ddot{\psi}$ for each of the segment types used in the path smoothing algorithm. The line segment is trivially defined as $\dot{\psi}=\ddot{\psi}=0$. The following subsections provide the derivation of expressions for $\dot{\psi}$ and $\ddot{\psi}$ for the remaining segment types.

\subsection{Arc}
The course angle of an arc segment is related linearly to the length
along the segment as
\begin{eqnarray}
\psi_a(s) & = & \psi_{0}+ks
\end{eqnarray}
where $k$ is the constant curvature of the arc segment. The derivative
of this with respect to path length is given by
\begin{eqnarray}
\psi_a^{\prime}(s) & = & k
\end{eqnarray}
and the second derivative is $\psi^{\prime\prime}=0.$ Then
the time derivatives of the course angle (see \eqref{eq:psidot} and \eqref{eq:psiddot}) are given by
\begin{eqnarray}
\dot{\psi}_a & = & k\dot{s} \label{eq:psidot_arc}\\
\ddot{\psi}_a & = & k\ddot{s}. \label{eq:psiddot_arc}
\end{eqnarray}

\subsection{Clothoid}
The course angle of the clothoid segment is given in \eqref{eq:clothoid_heading}.
The derivative with respect to the path length is given by
\begin{eqnarray*}
\psi^{\prime}_c(s) & = & k_{0}+\sigma_cs
\end{eqnarray*}
and the second derivative is given by
\begin{eqnarray*}
\psi^{\prime\prime}_c & = & \sigma_c.
\end{eqnarray*}
Then applying \eqref{eq:psidot} and \eqref{eq:psiddot} the
time derivatives are given by
\begin{eqnarray}
\dot{\psi}_c(s) & = & \left(k_{0}+\sigma_cs\right)\dot{s}\label{eq:psidot_clothoid}\\
\ddot{\psi}_c(s) & = & \sigma_c\dot{s}^{2}+\left(k_{0}+\sigma_cs\right)\ddot{s}.\label{eq:psiddot_clothoid}
\end{eqnarray}

\subsection{Fermat Spiral}
The course angle of a Fermat Spiral segment is given in \eqref{eq:fermat_course_angle} in terms of the polar angle $\theta$. In \cite{lekkas_continuous-curvature_2013}, a change of variables is presented as $u=\sqrt{\theta}$ which provides consistent sampling per path length. Then, the course angle can be expressed as
\begin{eqnarray*}
\psi(u) & = & u^{2}+\arctan\left(2u^{2}\right).
\end{eqnarray*}
This change of variables is useful because an expression for the time
derivative of $u$ exists and is given by
\begin{eqnarray*}
\dot{u} & = & \frac{\dot{s}}{c\sqrt{1+4u^{4}}}\\
 & = & \frac{\dot{s}\sqrt{4u^{4}+1}}{c\left(4u^{4}+1\right)}.
\end{eqnarray*}
Then the time derivative of the path segment can be computed as follows
\begin{eqnarray}
\dot{\psi} & = & \rho\frac{d\psi}{d u}\frac{d u}{d t}\nonumber \\
 & = & \rho\frac{2u\left(4u^{4}+3\right)}{4u^{4}+1}\dot{u} \label{eq:psidot_fermat}
\end{eqnarray}
where $\rho$ is the direction of the curve defined in  \eqref{eq:direction}.
The second derivative is computed by applying the quotient rule and
chain rule as
\begin{eqnarray*}
\ddot{\psi} & = & \frac{d}{dt}\left(\frac{2\dot{s}\left(4u^{5}+3u\right)\sqrt{4u^{4}+1}}{c\left(4u^{4}+1\right)^{2}}\right).
\end{eqnarray*}
The time derivative of the numerator is given by
\begin{eqnarray*}
\mathcal{\dot{N}} & = & 2\left[\ddot{s}\text{\ensuremath{\left(\left(4u^{5}+3u\right)\sqrt{4u^{4}+1}\right)}+\ensuremath{\dot{s}}}\mathcal{\mathcal{\dot{N}}}_{1}\right]
\end{eqnarray*}
where 
\begin{eqnarray*}
\mathcal{\dot{N}}_{1} & = & \frac{d}{dt}\left(\left(4u^{5}+3u\right)\sqrt{4u^{4}+1}\right)\\
 & = & \left[\left(20u^{4}+3\right)\sqrt{4u^{4}+1}+\frac{8u^{3}\left(4u^{5}+3u\right)}{\sqrt{4u^{4}+1}}\right]\dot{u}.
\end{eqnarray*}
The time derivative of the denominator is given by
\begin{eqnarray*}
\dot{\mathcal{D}} & = & 32c\left(4u^{4}+1\right)u^{3}\dot{u}
\end{eqnarray*}
then
\begin{eqnarray}
\ddot{\psi} & = & \rho\frac{\mathcal{\dot{N}\mathcal{D}-\mathcal{\dot{D}\mathcal{N}}}}{\mathcal{D}^{2}}.\label{eq:psiddot_fermat}
\end{eqnarray}

%% file: sections/Results.tex
This section provides results for the path smoothing and aircraft state generation. The first set of results will include a comparison of the corner smoothing techniques presented in Section \ref{sec:Path-Smoothing}. These results include a comparison between the path characteristics and the execution time for the methods presented. The second set of results show a comparison between the generated IMU signals. 
%The second set of results show the performance of the covariance propagation approach as it compares to a 6DOF UAV simulator with higher-fidelity acceleration and angular rates.

\subsection{Path Smoothing\label{subsec:PathResults}}
\begin{figure}
	\begin{centering}
		\includegraphics[width=0.85\columnwidth]{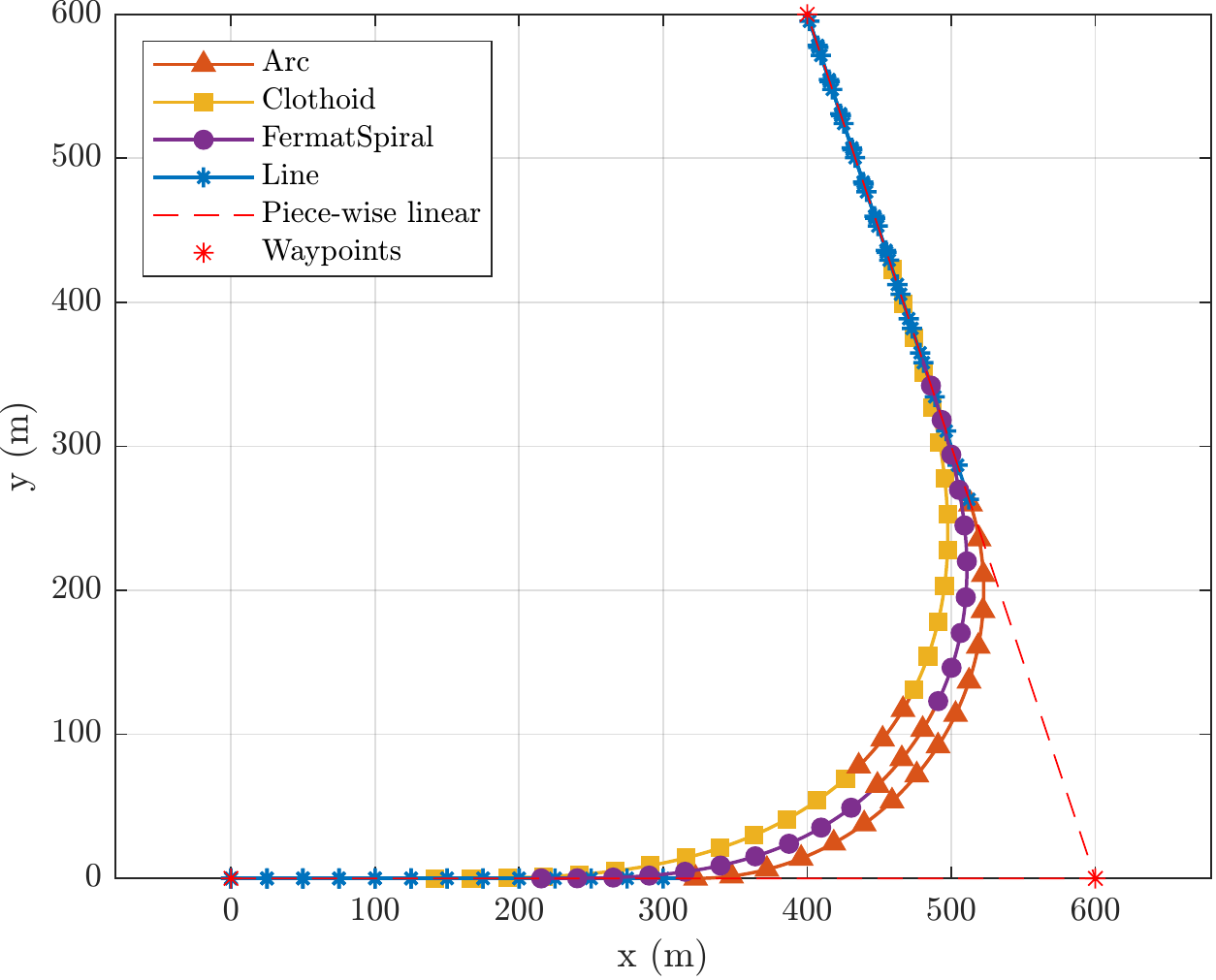}
		\par\end{centering}
	\caption{Continuous curvature path given 3 waypoints. Blue lines represent
		straight segments, yellow lines with square markers represent clothoid
		segments, purple lines with circle markers represent Fermat segments
		and the orange lines with triangle markers represents the $k_{max}$
		arc segment. The dashed red line indicates the piece-wise linear path
		and the red asterisk symbols represent the waypoints. \label{fig:res_3wps}}
\end{figure}

The algorithm defined in Section \ref{sec:Path-Smoothing}
was implemented in Matlab where a series of waypoints could be converted
to a continuous curvature path. Fig. \ref{fig:res_3wps} shows a
piece-wise linear path smoothed using the three methods presented
here. This was generated
using \textbf{$k_{max}=2.1$, }and $k_{max}^{\prime}=3$.

\begin{figure}
	\begin{centering}
		\includegraphics[width=1\columnwidth]{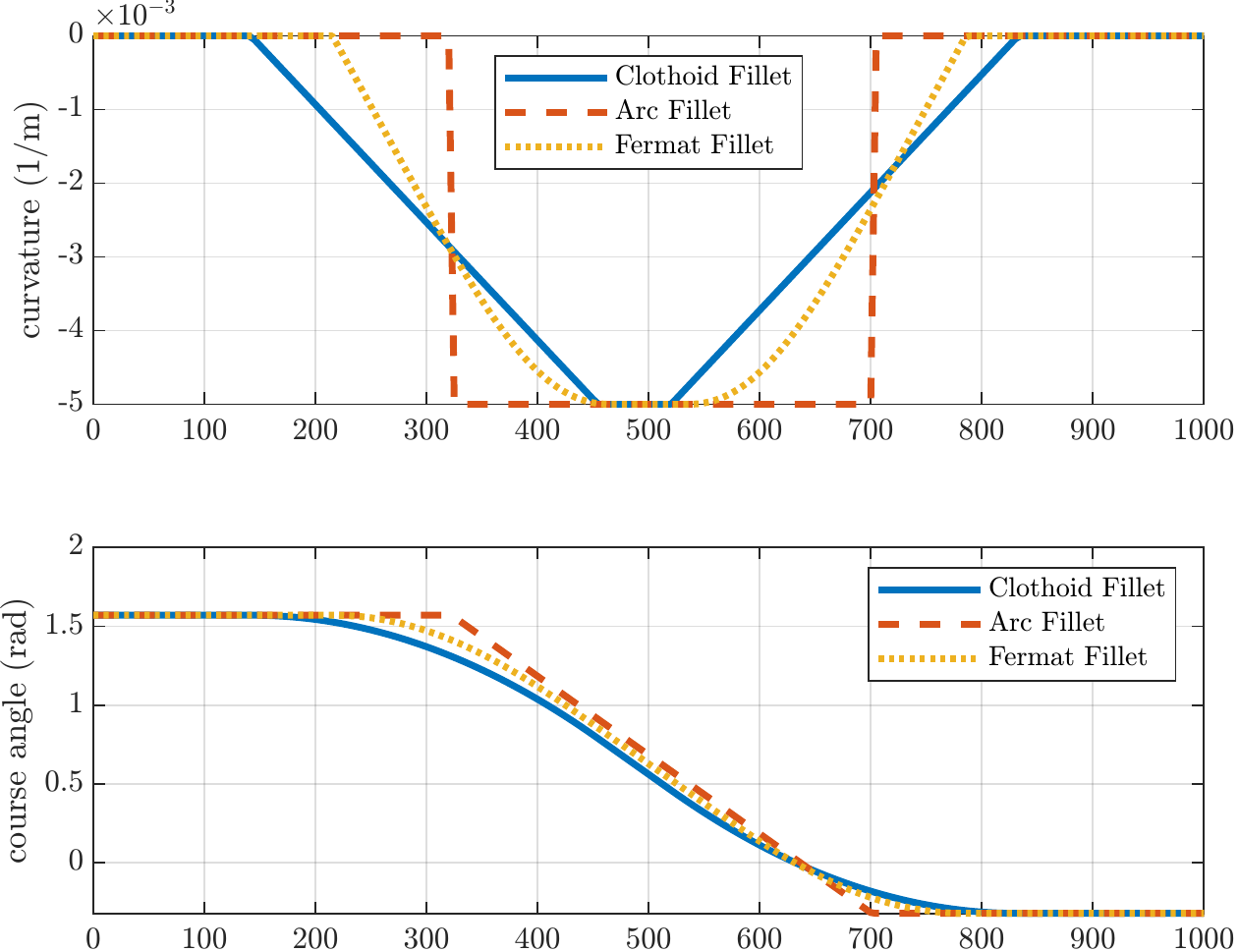}
		\par\end{centering}
	\caption{Curvature and course angle of the three path smoothing methods. Clothoid
		fillet (solid), arc fillet (dashed), Fermat fillet (dotted).\label{fig:Curvature-and-course}}
\end{figure}

The course angle and curvature for the scenario shown in Fig. \ref{fig:res_3wps}
is shown in Fig. \ref{fig:Curvature-and-course}. For this scenario
all three methods achieve the maximum curvature but the differences
in the shape for the curvature and course angle are apparent. The
curvature plot shows a step change for the Arc Fillet method, a linear
change for the Clothoid Fillet method, and a smooth transition
for the Fermat Fillet method.

The shape of the curvature and course angle graphs should be considered when deciding what corner smoothing algorithm to use. For example, the arc fillet method has a step change in curvature which is infeasible for the fixed-wing aircraft application presented in this work. This highlights the need to consider the dynamics of the target system and the desired maneuvers when selecting a corner smoothing algorithm.

The processing time for the 3 methods were compared using
a series of 12 waypoints. The Clothoid Fillet method is the slowest (due to the Fresnel
Integrals), and the Arc Fillet method is the fastest. The Fermat Fillet
method provides a $63.3\%$ improvement in execution time when compared
with the Clothoid Fillet method, whereas the Arc Fillet method provides
an $87.98\%$ improvement. These results provide a framework for deciding
which method best suits the target application. For example, if the
application does not require continuous curvature, the Arc Fillet
method is very efficient and fulfills the objective. However, if continuous
curvature is desired, the Fermat Fillet or Clothoid Fillet methods
should be considered and if execution time is a primary concern, the preference
should be given to the Fermat Fillet method.

% \begin{table}
% 	\caption{Execution time for fillet methods for a series of 12 waypoints \label{tab:method_execution_time}}
% 	\begin{centering}
% 	\def\arraystretch{1.5} % Add vertical buffer to table for readability
% 		\begin{tabular}{c|c}
% %			\hline 
% 			\textbf{Method} & \textbf{Execution Time (s)} \tabularnewline
% %			\hline 
% 			\hline 
% 			\textbf{Clothoid Fillet} & 0.1397\tabularnewline
% 			\hline 
% 			\textbf{Fermat Fillet} & 0.0512\tabularnewline
% 			\hline 
% 			\textbf{Arc Fillet} & 0.0168\tabularnewline
% %			\hline 
% 		\end{tabular}
% 		\par\end{centering}
% \end{table}

\subsection{State Derivative Results \label{subsec:StateDerivRes}}
The previous subsection showed results for the three path smoothing methods presented in this work. This section will extend these results and present the specific force and angular rates generated using the IMU signal generation method shown in Section \ref{sec:State-Generation}. Due to the discontinuities in course angle for the Arc Fillet method, it will be omitted from this analysis.

\begin{figure}
	\noindent\begin{minipage}[t]{1\columnwidth}%
		\begin{center}
			\scalebox{0.8}{\input{figures/ASG_results_block_diagram.tex}}
			\par\end{center}%
	\end{minipage}
	\caption{Block diagram of ASG and RII used for the results comparison. \label{fig:results_block_diagram}}
\end{figure}
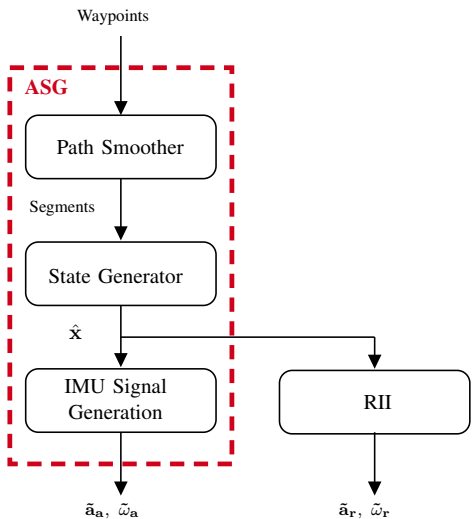

\begin{figure}
	\begin{centering}
		\includegraphics[width=0.85\columnwidth]{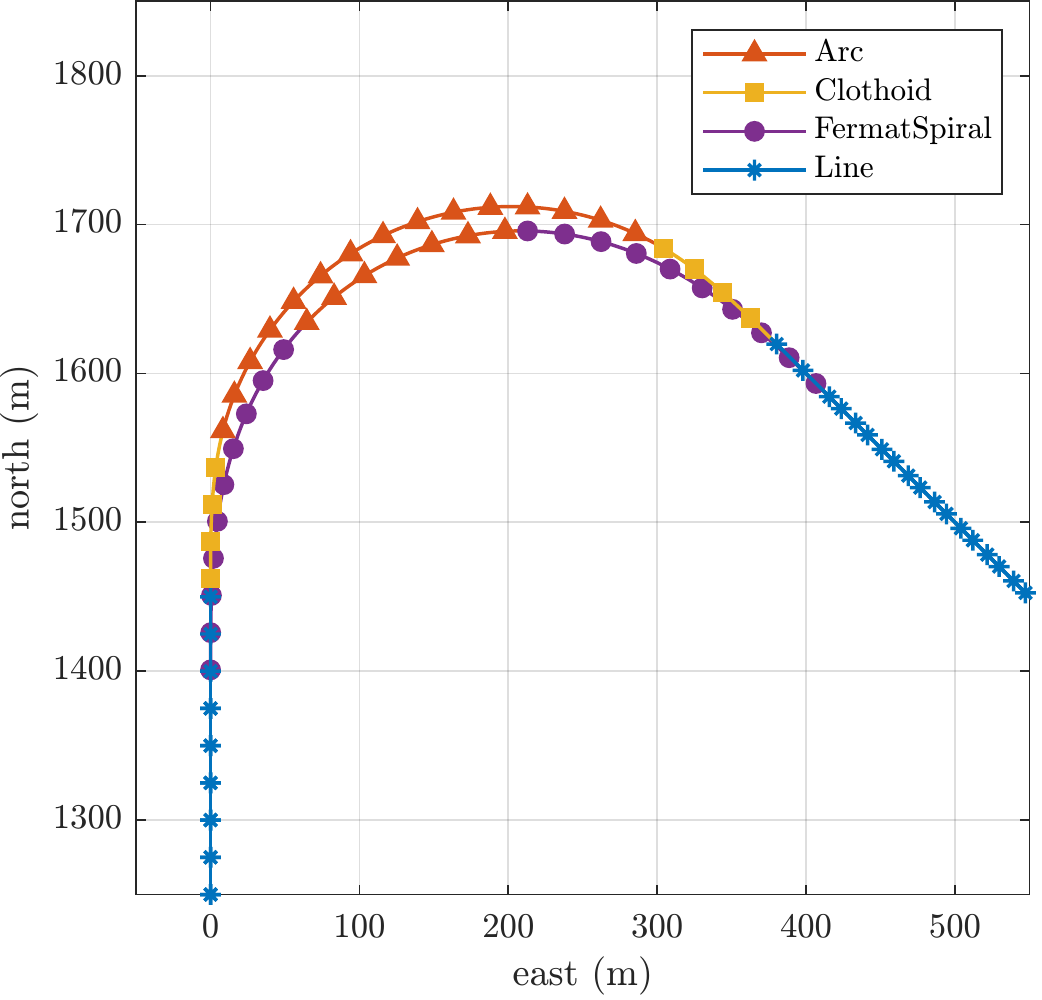} 
		\par
	\end{centering}
	\caption{First turn of nominal trajectory scenario showing the Clothoid Fillet and Fermat Fillet paths. The paths were created with $k_{max} = 0.005\ 1/rad$, and $k^\prime_{max} = 0.00005\ 1/rad/m$. \label{fig:state_deriv_2d}}
\end{figure}

The IMU signals generated using the method shown in Section \ref{sec:State-Generation} will be compared with the RII method presented in Section \ref{subsec:RII}. Fig. \ref{fig:results_block_diagram} shows a diagram of how the accelerometer, $\boldsymbol{\tilde{a}} = [a_x\ a_y\ a_z]^\intercal$, and gyro, $\boldsymbol{\tilde{\omega}} = [g_x\ g_y\ g_z]^\intercal$, measurements are generated using the ASG and RII methods. Two simple scenarios of four waypoints (two turns) were used to compare the methods. The first method will be referred to as the nominal trajectory scenario with $k_{max} = 0.005\ 1/rad$, and $k^\prime_{max} = 0.00005\ 1/rad/m$. The second will be referred to as the aggressive trajectory scenario that has waypoints closer together and with $k_{max} = 0.01\ 1/rad$, and $k^\prime_{max} = 0.0002\ 1/rad/m$. Fig. \ref{fig:state_deriv_2d} shows one of the turns in the nominal trajectory scenario with the Clothoid Fillet and Fermat Fillet methods. 

\begin{figure}
	\begin{centering}
		\includegraphics[width=1\columnwidth]{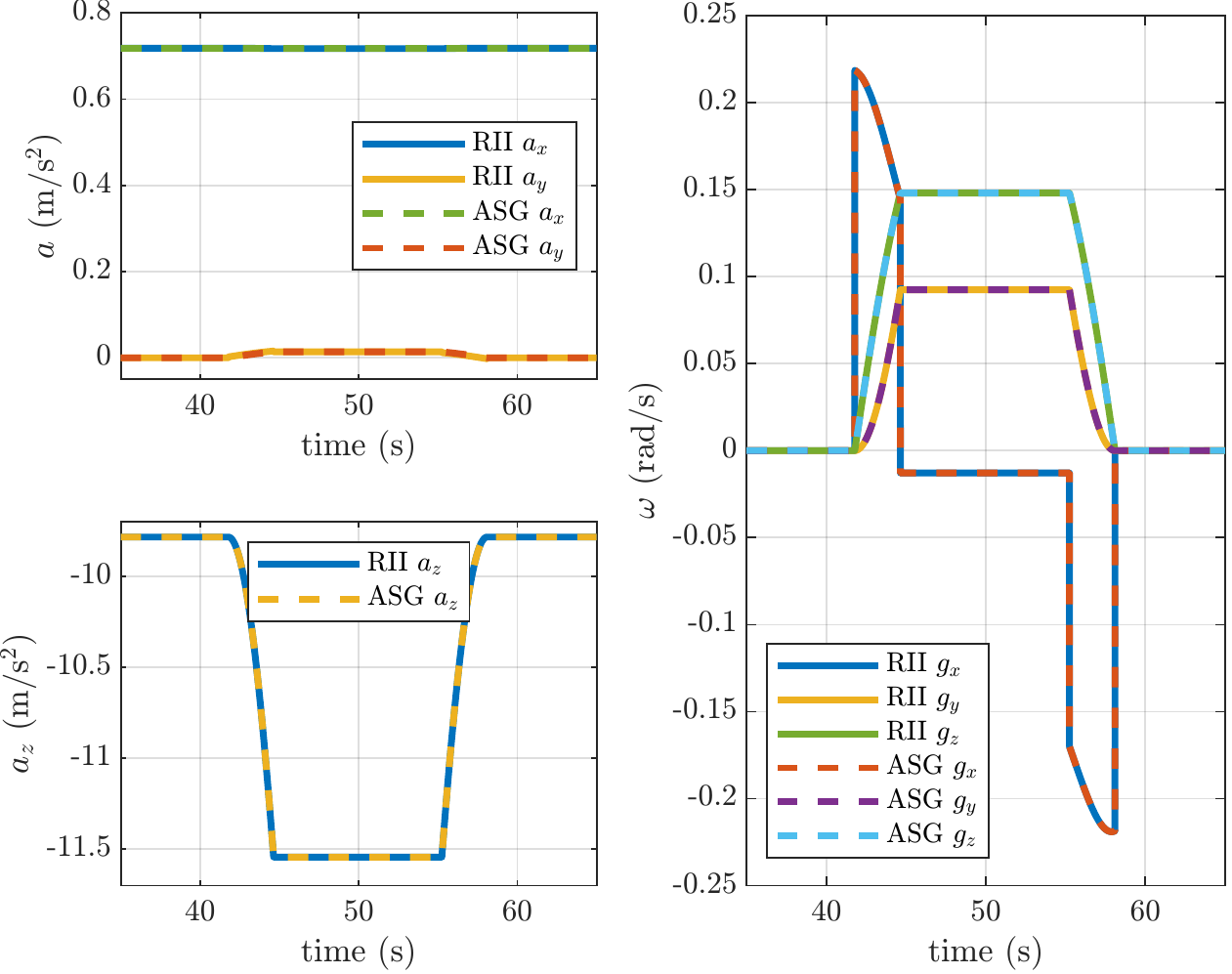}
		\par
	\end{centering}
	\caption{Accelerometer and gyro measurements for the nominal trajectory scenario using the Clothoid Fillet corner smoother. \label{fig:state_deriv_cloth}}
\end{figure}

\begin{figure}
	\begin{centering}
		\includegraphics[width=1\columnwidth]{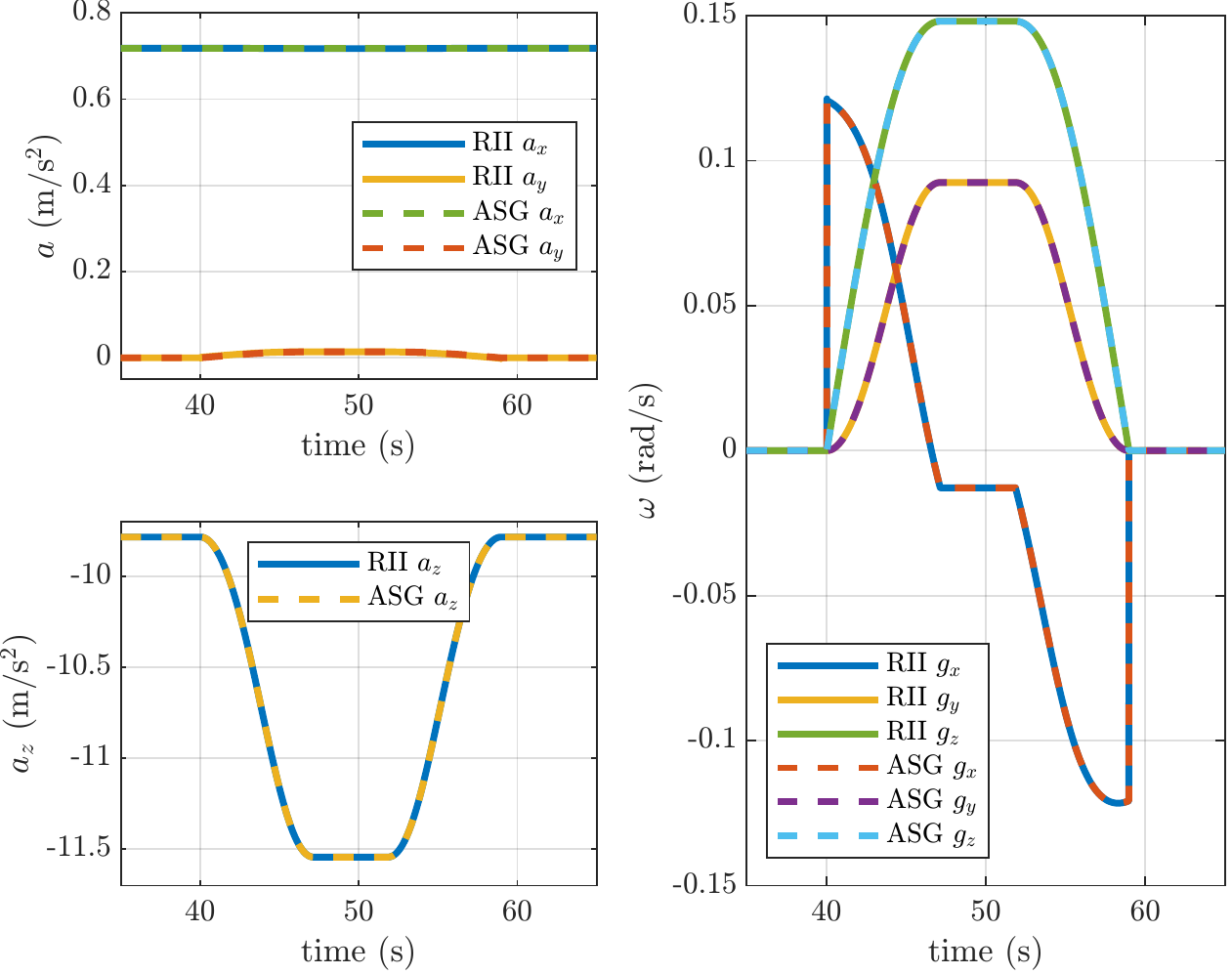}
		\par
	\end{centering}
	\caption{Accelerometer and gyro measurements for the nominal trajectory scenario using the Fermat Fillet corner smoother. \label{fig:state_deriv_fermat}}
\end{figure}

The accelerometer and gyro measurements for the Clothoid and Fermat Fillet methods for the nominal trajectory scenario are provided in Figs. \ref{fig:state_deriv_cloth} and \ref{fig:state_deriv_fermat}, respectively. The IMU signals for both methods match the baseline results well. The slight differences between the RII method and ASG method are difficult to see in Figs. \ref{fig:state_deriv_cloth} and \ref{fig:state_deriv_fermat} so an error metric will be introduced in the following paragraphs to illustrate the differences.

As discussed in Section \ref{subsec:RII}, the RII method relies on a discrete derivative taken at each time step. For a small time step, $dt$, this method provides consistent results. However, in some applications, accurate IMU signals are desired with larger time steps. This can be particularly useful in mission and path planning scenarios where reducing processor usage is required. One benefit to the ASG method is that the IMU signal generation avoids the derivative and maintains accuracy for large time steps.

To quantify the effect of increasing the time step on these two methods, baseline IMU signals were generated using the RII method with a small time step ($dt = 0.001s$). IMU signals for time steps ranging from $dt = 0.1s$ to $dt = 3.0s$ were generated using both methods. 
% and an error value, $\Upsilon$, was calculated for each time step. Where
% \begin{equation}
% \Upsilon_x = \sum_{\tilde{t}=0}^{\tilde{t}_f} ||\boldsymbol{\hat{x}_r}(\hat{t}_k) - \boldsymbol{\tilde{x}_a}(\tilde{t}_k)||_2 dt \label{eq:upsilon}
% \end{equation}
% The decorated time symbols are used to indicate that though the baseline IMU signal will have more samples than the error signal, the subtraction will compare signals at the same time during the simulation so $\hat{t}_k=\tilde{t}_k$. The $x$ subscript can be replaced $a$ or $\omega$ if the error is being calculated for the accelerometer or gyro signals, respectively. 

\begin{figure}
	\begin{centering}
		\includegraphics[width=1\columnwidth]{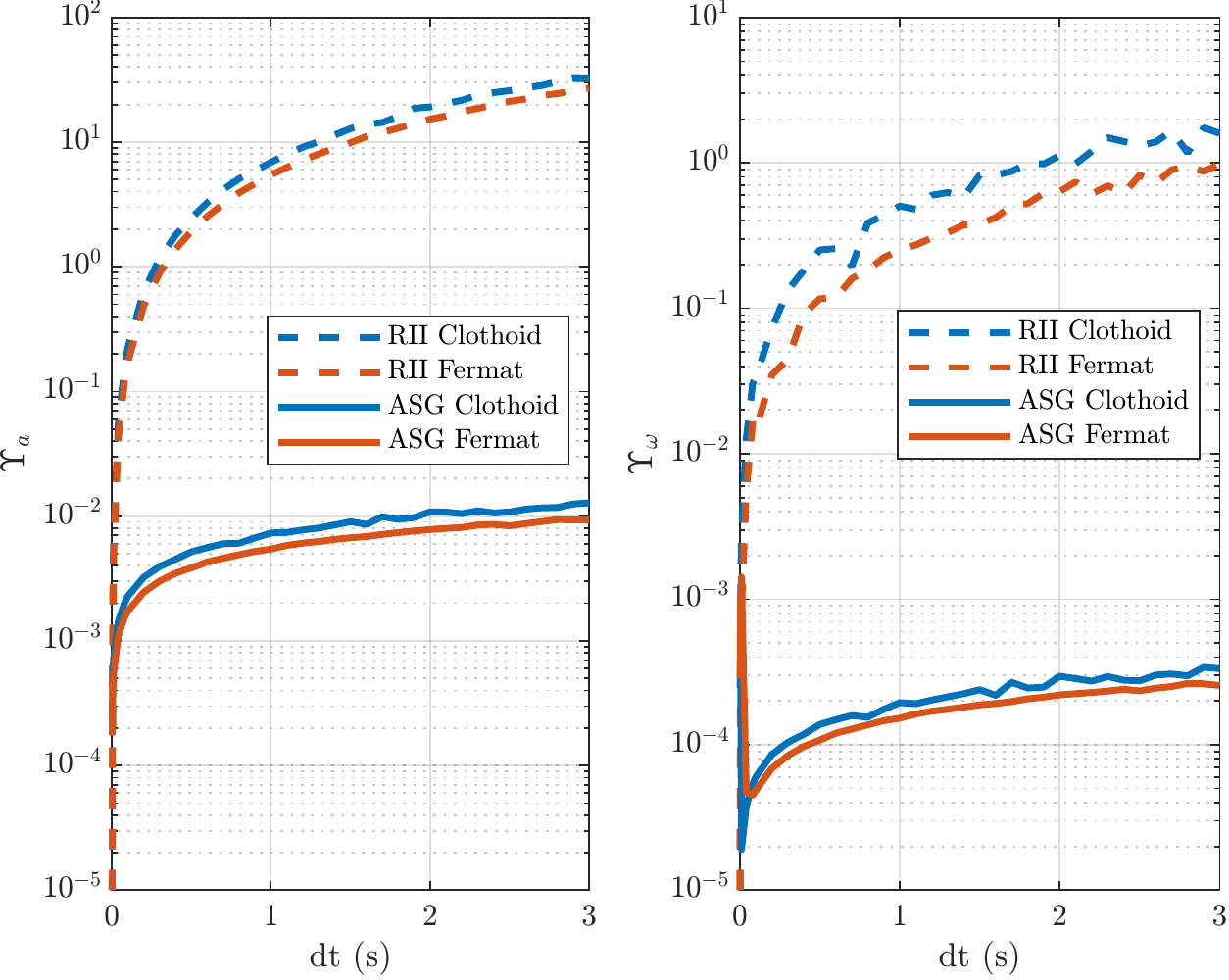} 
		\par
	\end{centering}
	\caption{Integrated norm of the error for the accelerometer (left) and gyro (right) measurements for the nominal trajectory scenario for a range of sample times. The plots show errors of both IMU signal generation methods (RII, ASG) for both Clothoid Fillet (C) and Fermat Fillet (F) corner smoothing algorithms. The y-axis was plotted with a log-scale to more clearly show the differences. \label{fig:state_deriv_error}}
\end{figure}

The integrated norm of the error, $\Upsilon$, of the IMU signals for the nominal trajectory scenario is shown in Fig. \ref{fig:state_deriv_error}. The errors associated with the RII method grow more quickly than the ASG method as the step size increases for both the accelerometer and gyro measurements.The ASG method results in nearly constant errors for both corner smoothing methods. The ASG method clearly provides more accurate IMU signals for the time steps shown.

\begin{figure}
	\begin{centering}
		\includegraphics[width=1\columnwidth]{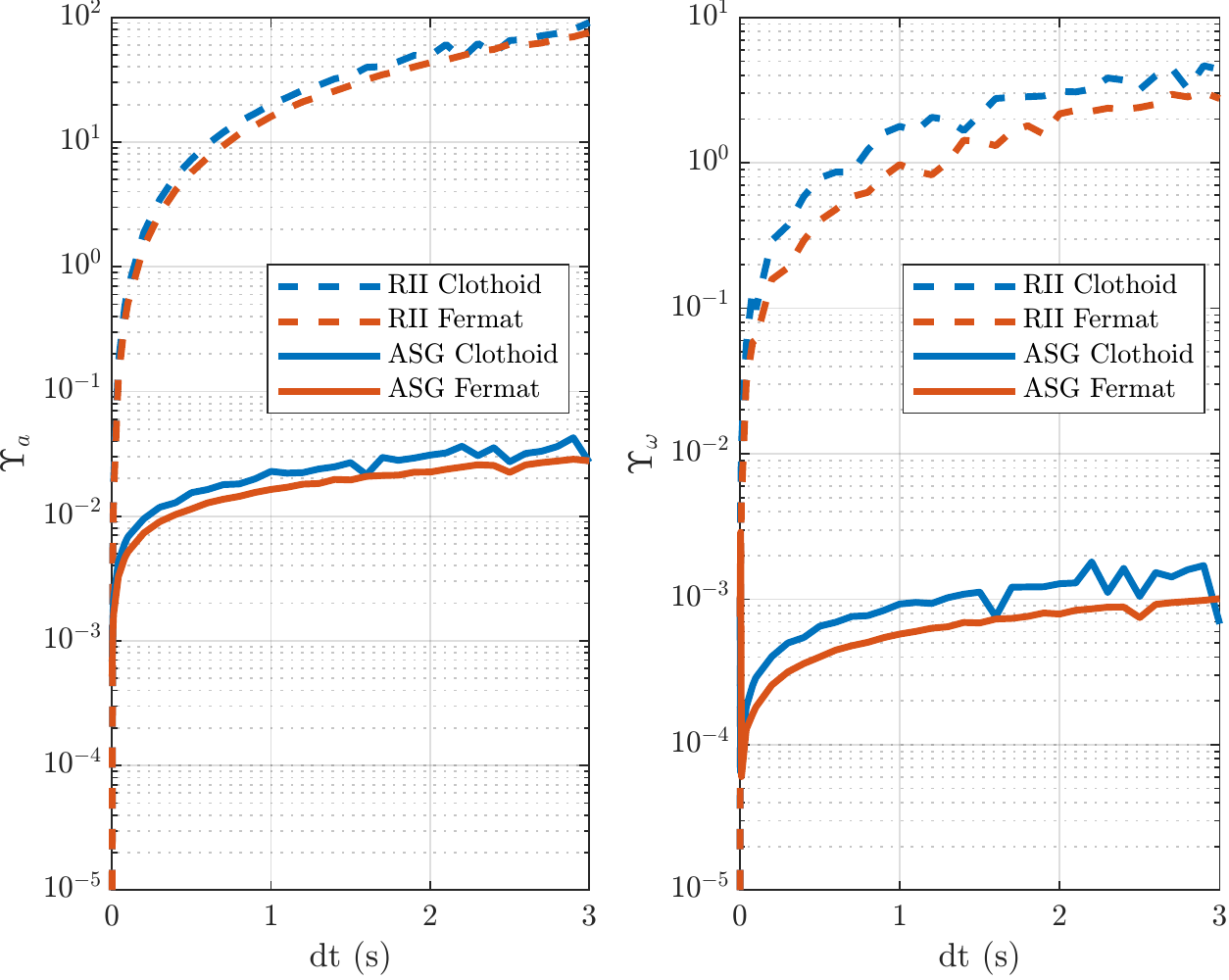}
		\par
	\end{centering}
	\caption{Integrated norm of the error for the accelerometer (left) and gyro (right) measurements for the aggressive trajectory scenario for a range of sample times. The plots show errors of both IMU signal generation methods (RII, ASG) for both Clothoid Fillet (C) and Fermat Fillet (F) corner smoothing algorithms. The y-axis was plotted with a log-scale to more clearly show the differences. \label{fig:state_deriv_error_2}}
\end{figure}

The integrated norm of the error, $\Upsilon$, for the aggressive trajectory scenario is shown in Fig. \ref{fig:state_deriv_error_2}. This trajectory has waypoints that are closer together and $k_{max} = 0.01\ 1/rad$, and $k^\prime_{max} = 0.0002\ 1/rad/m$. The error plots indicate that the ASG method out performs the RII method for this trajectory as well, and the general error characteristics are consistent with the nominal trajectory scenario. It should be observed that the errors are higher for the aggressive trajectory scenario than the nominal trajectory scenario. 

%% file: figures/ASG_results_block_diagram.tex
\tikzset{every picture/.style={line width=0.75pt}} %set default line width to 0.75pt        

\begin{tikzpicture}[x=0.75pt,y=0.75pt,yscale=-1,xscale=1]
%uncomment if require: \path (0,418); %set diagram left start at 0, and has height of 418

%Rounded Rect [id:dp13705846124268173] 
\draw   (240,88) .. controls (240,83.58) and (243.58,80) .. (248,80) -- (352,80) .. controls (356.42,80) and (360,83.58) .. (360,88) -- (360,112) .. controls (360,116.42) and (356.42,120) .. (352,120) -- (248,120) .. controls (243.58,120) and (240,116.42) .. (240,112) -- cycle ;
%Straight Lines [id:da19733012194326882] 
\draw    (300,30) -- (300,77) ;
\draw [shift={(300,80)}, rotate = 270] [fill={rgb, 255:red, 0; green, 0; blue, 0 }  ][line width=0.08]  [draw opacity=0] (8.93,-4.29) -- (0,0) -- (8.93,4.29) -- cycle    ;
%Rounded Rect [id:dp05166410716139902] 
\draw   (240,168) .. controls (240,163.58) and (243.58,160) .. (248,160) -- (352,160) .. controls (356.42,160) and (360,163.58) .. (360,168) -- (360,192) .. controls (360,196.42) and (356.42,200) .. (352,200) -- (248,200) .. controls (243.58,200) and (240,196.42) .. (240,192) -- cycle ;
%Straight Lines [id:da643161928058195] 
\draw    (300,120) -- (300,156) ;
\draw [shift={(300,159)}, rotate = 270] [fill={rgb, 255:red, 0; green, 0; blue, 0 }  ][line width=0.08]  [draw opacity=0] (8.93,-4.29) -- (0,0) -- (8.93,4.29) -- cycle    ;
%Rounded Rect [id:dp6635067447596747] 
\draw   (400,249) .. controls (400,244.58) and (403.58,241) .. (408,241) -- (512,241) .. controls (516.42,241) and (520,244.58) .. (520,249) -- (520,273) .. controls (520,277.42) and (516.42,281) .. (512,281) -- (408,281) .. controls (403.58,281) and (400,277.42) .. (400,273) -- cycle ;
%Rounded Rect [id:dp01623461178291863] 
\draw   (240,248) .. controls (240,243.58) and (243.58,240) .. (248,240) -- (352,240) .. controls (356.42,240) and (360,243.58) .. (360,248) -- (360,272) .. controls (360,276.42) and (356.42,280) .. (352,280) -- (248,280) .. controls (243.58,280) and (240,276.42) .. (240,272) -- cycle ;
%Straight Lines [id:da5415722541905053] 
\draw    (300,200) -- (300,236) ;
\draw [shift={(300,239)}, rotate = 270] [fill={rgb, 255:red, 0; green, 0; blue, 0 }  ][line width=0.08]  [draw opacity=0] (8.93,-4.29) -- (0,0) -- (8.93,4.29) -- cycle    ;
%Shape: Rectangle [id:dp4338657231190006] 
\draw  [color={rgb, 255:red, 208; green, 2; blue, 27 }  ,draw opacity=1 ][dash pattern={on 6.75pt off 4.5pt}][line width=2.25]  (230,50) -- (370,50) -- (370,300) -- (230,300) -- cycle ;
%Straight Lines [id:da10838420748373978] 
\draw    (460,220) -- (460,237) ;
\draw [shift={(460,240)}, rotate = 270] [fill={rgb, 255:red, 0; green, 0; blue, 0 }  ][line width=0.08]  [draw opacity=0] (8.93,-4.29) -- (0,0) -- (8.93,4.29) -- cycle    ;
%Straight Lines [id:da22322794629288567] 
\draw    (300,220) -- (460,220) ;
%Straight Lines [id:da17194671999291766] 
\draw    (460,280) -- (460,317) ;
\draw [shift={(460,320)}, rotate = 270] [fill={rgb, 255:red, 0; green, 0; blue, 0 }  ][line width=0.08]  [draw opacity=0] (8.93,-4.29) -- (0,0) -- (8.93,4.29) -- cycle    ;
%Straight Lines [id:da7169854138925853] 
\draw    (300,280) -- (300,317) ;
\draw [shift={(300,320)}, rotate = 270] [fill={rgb, 255:red, 0; green, 0; blue, 0 }  ][line width=0.08]  [draw opacity=0] (8.93,-4.29) -- (0,0) -- (8.93,4.29) -- cycle    ;

% Text Node
\draw (271,12) node [anchor=north west][inner sep=0.75pt]  [font=\footnotesize] [align=left] {Waypoints};
% Text Node
\draw (251,94) node [anchor=north west][inner sep=0.75pt]   [align=left] {\begin{minipage}[lt]{70.76216000000001pt}\setlength\topsep{0pt}
\begin{center}
Path Smoother
\end{center}

\end{minipage}};
% Text Node
\draw (245,175) node [anchor=north west][inner sep=0.75pt]   [align=left] {\begin{minipage}[lt]{75.29776000000001pt}\setlength\topsep{0pt}
\begin{center}
State Generator
\end{center}

\end{minipage}};
% Text Node
\draw (238,56) node [anchor=north west][inner sep=0.75pt]  [color={rgb, 255:red, 208; green, 2; blue, 27 }  ,opacity=1 ] [align=left] {\textbf{ASG}};
% Text Node
\draw (242,245) node [anchor=north west][inner sep=0.75pt]   [align=left] {\begin{minipage}[lt]{79.82112000000001pt}\setlength\topsep{0pt}
\begin{center}
IMU Signal Generation
\end{center}

\end{minipage}};
% Text Node
\draw (450,255) node [anchor=north west][inner sep=0.75pt]   [align=left] {\begin{minipage}[lt]{15.755328pt}\setlength\topsep{0pt}
\begin{center}
RII
\end{center}

\end{minipage}};
% Text Node
\draw (266,210.4) node [anchor=north west][inner sep=0.75pt]    {$\hat{\mathbf{x}}$};
% Text Node
\draw (241,132) node [anchor=north west][inner sep=0.75pt]   [align=left] {{\footnotesize Segments}};
% Text Node
\draw (436,323.4) node [anchor=north west][inner sep=0.75pt]  [font=\footnotesize]  {$\mathbf{\tilde{a}_{r}} ,\ \mathbf{\tilde{\omega }_{r}}$};
% Text Node
\draw (276,323.4) node [anchor=north west][inner sep=0.75pt]  [font=\footnotesize]  {$\mathbf{\tilde{a}_{a}} ,\ \mathbf{\tilde{\omega }_{a}}$};

\end{tikzpicture}

%% file: sections/Conclusion.tex
This work has presented ASG as a method for generating aircraft states and IMU signals along a candidate path. The method includes a corner smoothing algorithm to convert a series of waypoints into a smoothed reference trajectory and uses path segment geometry to generate the aircraft states. The final stage of the method applies curvilear motion theory and vehicle maneuver dynamics to generate IMU signals along the candidate path. The generated states and IMU signals can be applied to a variety of applications, including general path planning, and covariance propagation for inertial navigation systems.

Three corner smoothing approaches were presented in a unified framework. The Clothoid Fillet and Fermat Fillet algorithms can be parameterized to respect vehicle constraints in curvature and curvature rate. The results showed that the Fermat Fillet requires less processing time than the Clothoid Fillet due to the Fresnel integrals. The Arc Fillet method requires the least processing time but has discontinuities in curvature and is thus not suitable for a fixed wing aircraft.

The results for the IMU signal generation stage show that the ASG method performs better than the RII method by avoiding a discrete derivative. This allows for large step sizes in sampling when using the ASG method.